\documentclass[reprint,onecolumn,aps,byrevtex,showpacs,nofootinbib,notitlepage]{revtex4-1}
\pdfoutput=0
\usepackage{graphicx}
\usepackage{psfrag}
\usepackage{amsmath}
\usepackage{amssymb}
\usepackage{mathtools}
\usepackage{xcolor}
\usepackage{comment}
\usepackage{subfig}
\usepackage{caption}
\usepackage{geometry}
\newcommand{\be}{\begin{equation}}
\newcommand{\ee}{\end{equation}}
\newcommand{\ra}{\rightarrow}

\def\bc{\begin{center}}
\def\ec{\end{center}}
\def\bea{\begin{eqnarray}}
\def\eea{\end{eqnarray}}

\begin{document}

\title{A large deviation perspective on ratio observables in reset processes: robustness of rate functions}

\author{Francesco Coghi}
\email{f.coghi@qmul.ac.uk}
\affiliation{School of Mathematical Sciences, Queen Mary University of London, London E1 4NS, UK}
\author{Rosemary J. Harris}
\affiliation{School of Mathematical Sciences, Queen Mary University of London, London E1 4NS, UK}

\date{\today}

\begin{abstract}
We study large deviations of a ratio observable in discrete-time reset processes. The ratio takes the form of a current divided by the number of reset steps and as such it is not extensive in time. A large deviation rate function can be derived for this observable via contraction from the joint probability density function of current and number of reset steps. The ratio rate function is differentiable and we argue that its qualitative shape is `robust', i.e. it is generic for reset processes regardless of whether they have short- or long-range correlations. We discuss similarities and differences with the rate function of the efficiency in stochastic thermodynamics.

\end{abstract}

\maketitle


\section{Introduction}
\label{sec:intro}

Stochastic reset processes have the property of being re-initialised at random times to a specific initial condition, which can be a particular probability distribution, or a fixed state. Although their natural framework is the mathematical language of renewal theory, see \cite{grimmett2001probability,feller2008introduction}, in the last decade reset processes have also been widely studied in the statistical physics community. They have been used to model the population dynamics after catastrophic events \cite{brockwell1985extinction,kyriakidis1994stationary,di2003m}, the dynamics of queues \cite{glynn1994large,di2003m}, the path of a transport protein moving on a cytoskeleton filament \cite{meylahn2015biofilament,meylahn2015large}, and also the foraging of animals in nature \cite{benichou2011intermittent}. Clearly, different observables are of interest in various real world scenarios, for instance, mean first passage times in search strategies for animal foraging \cite{evans2011diffusion}, or additive functionals of time (e.g. position or current) in cellular transport proteins. Reset applications are not only restricted to classical environments, but can also be extended to quantum mechanical systems \cite{gherardini2016stochastic,mukherjee2018quantum,rose2018spectral}.

In this paper, we focus attention on a different class of observables: ratios between addictive functionals of time. In finance, for example, one can calculate the Sharpe ratio, which gives a good estimate of the excess expected return of an investment given its volatility \cite{lo2002statistics}. 
In stochastic thermodynamics, physicists have recently studied thermodynamic/kinetic uncertainty relations, which give bounds for a type of ratio observable \cite{harris2019thermodynamic,di2018kinetic,barato2018unifying}. In the same field, there are studies of fluctuations of the efficiency, defined as the ratio between the output work and the input heat, of small-scale engines working in an energetically unstable environment \cite{verley2014unlikely,verley2014universal,gingrich2014efficiency,proesmans2015stochastic,polettini2015efficiency,vroylandt2019efficiency}. The latter is relevant in biology, where understanding the efficiency of molecular motors \cite{julicher1997modeling}, e.g. myosin heads moving on actin filaments \cite{kitamura1999single,cyranoski2000swimming}, is important in medical applications. Ratios also appear in probability theory, for example representing maximum likelihood estimators for Ornstein-Uhlenbeck processes with and without shift \cite{bercu2002sharp,bercu2015large}.

Of particular significance for the present work, it is argued in stochastic thermodynamics \cite{verley2014unlikely,verley2014universal} that the fluctuating efficiency can be described by a universal large deviation rate function shape, characterised by having a maximum (as well as the usual minimum) and tails tending to a horizontal asymptote. These intriguing features have attracted considerable recent attention with physical explanations proposed for their appearance \cite{verley2014unlikely,verley2014universal,gingrich2014efficiency}. Understanding both typical efficiency values, attained in the long-time limit, and fluctuations, arising in finite time, is important for predicting the performance of nano-motors, which can now be realised experimentally \cite{martinez2016brownian}. Beyond this practical example, uncovering the general features of ratio quantities contributes to building the theoretical framework of non-equilibrium statistical mechanics where dynamical fluctuations, studied by means of large deviation theory, play a crucial role. In this spirit, we present here a rigorous analysis of ratio observables associated to a particular class of stochastic processes: although such ratios are not true efficiencies, they share many features, e.g. the tail shape, and thus help to elucidate the underlying mathematical structure.

We now outline the concrete details of our approach. In this paper we study the ratio of the integrated current and the number of resets in a discrete-time reset process, aiming to understand its probability measure in the exponential scaling limit by means of large deviation theory. We prove that a large deviation principle is valid for this quantity by means of a contraction principle allowing us to transfer, under a continuous mapping, the large deviation principle that holds for the joint observable $($current, number of reset steps$)$ to the ratio observable. We then investigate the form of the obtained large deviation rate function in several situations, and we notice in all cases that, although tails are always bounded from above by a horizontal asymptote, the characteristic fluctuating efficiency maximum \cite{verley2014unlikely,verley2014universal,gingrich2014efficiency} is not present. Indeed, while the asymptotic shape is a notable property of ratio observables, which often present heavy tails and lack of exponential tightness in their distributions, the maximum can be thought as a geometric consequence of having both positive and negative 
fluctuations in the denominator. The main result we find relates to the `robustness' of the large deviation ratio rate function. We argue that the qualitative shape of the rate function is generic for reset processes whether they have short- or long-range correlations. In particular, we show that the rate function is differentiable. In contrast, we prove (calculation in the appendix) that when the reset nature of the process is lost, i.e. the numerator of the ratio observable is independent from the denominator, a `mode-switching' phase transition in the fluctuations of the ratio appears, and the rate function is not differentiable.

\section{Models and methods}
\label{sec:methods}

\subsection{Model framework}
\label{subsec:simplmod}

The reset process we consider has the property of being returned at random times to a certain initial condition represented by a `fixed' internal state. For our purposes, it suffices to think of a discrete-time random walk with hopping probabilities that depend on the time since reset. Here, the reset can be thought as restarting a clock variable which controls the dynamics.

In describing our models we find it useful to split the reset process into two layers. The bottom layer is a discrete-time stochastic process $\mathbf{X}_n = \left( X_1,X_2,...,X_n \right)$ composed of $n$ Bernoulli random variables of parameter $r$, i.e. with probability $r$, $X_i=1$ (corresponding to a reset at the $i$-th time step), otherwise $X_i=0$ (no reset). The top layer is a discrete-time (but continuous-space) random walk $\mathbf{Y}_n = \left( Y_1,Y_2,...,Y_n \right)$, taking a jump, at the $i$-th time step, $Y_i$ according to a certain probability function depending in general on the time since the last reset. For definiteness, we think of periodic boundary conditions since we are chiefly interested in the net movement of the random walker rather than its position. We refer to the bottom layer as the on-off process, and to the top layer as the random walk. The reset nature of the process arises from the restarting of the internal clock (happening when $X_i=1$), re-initialising the dynamical rules for the movement of the random walk in the top layer. 

In this framework the observables we study are: the empirical number of reset steps, the empirical current, and their ratio. They read respectively
\begin{align}
N_n &= \sum_{i=1}^n X_i , \\
J_n &= \sum_{i=1}^n Y_i , \\
\Omega_n &= \frac{J_n}{N_n} .
\end{align}
We focus on the long-time behaviour of these observables, with the aim of studying the exponential scaling of the ratio probability density function. The intensive (rescaled) observables are: $N_n/n \coloneqq \eta \in D \triangleq [0,1]$, $J_n/n \coloneqq j \in \mathbb{R}$, and $\Omega_n \coloneqq \omega \in \mathbb{R}$. Note that $N_n$, the denominator of $\Omega_n$, can take only positive values with $0$ included. The possible divergence in the ratio will be important later when considering the validity of the so-called contraction principle.

The reset character arises from the correlations between $N_n$ and $J_n$ which come from two sources. Firstly, we typically enforce $Y_i = 0$ when $X_i = 1$ (corresponding to freezing of the current during reset in the spirit of \cite{harris2017phase}). Secondly, we allow the possibility that the distribution of $Y_i$ when $X_i=0$ depends on the time elapsed since the last reset (i.e. the internal clock time). It is the presence of these correlations that makes our study of reset processes a difficult, and interesting, task. To gain some initial intuition and to demonstrate the mathematical techniques we first introduce two minimal models where correlations are minimised. Later we will consider models with both types of correlations discussed above, as well as those where the on-off process is itself correlated.

The first minimal model, called $M_1$, does not present any of these correlations, i.e. it is characterised by having completely uncorrelated layers. To be more specific, regardless of what happens in the bottom layer, the random walk in the top layer takes a jump at time step $i$ according to a Gaussian distribution of mean $\mu$ and variance $\sigma^2 = 2$. 
The second minimal model, called $M_2$, presents the first kind of correlations introduced above; it is a type of `lazy' random walk as in \cite{harris2017phase}. In contrast to $M_1$, now the top layer is coupled with the bottom one -- the random walk takes a jump at the $i$-th time step only if a reset does not happen in the other layer ($X_i = 0$), according to the Gaussian probability density function introduced above.

\subsection{Large deviation principles for the empirical means}
\label{subsec:LDPmeans}

A large deviation principle (LDP) holds for a particular (continuous) observable $A_n$ associated to a stochastic process if  
\be
\label{eq:ldp}
\lim_{n \ra \infty} \frac{1}{n} \ln \mathbb{P} \left( \frac{A_n}{n} \in [a,a+da] \right) = I(a) ,
\ee
where $I \in [0, \infty)$ is the so-called large deviation rate function. In our convention $\mathbb{P}$ is a probability measure, whereas $P$ is a probability density function, i.e. $\mathbb{P}(A_n/n \in [a,a+da]) = P(a)da$. With a little abuse of notation, the shorthand $\mathbb{P}(A_n/n=a) \coloneqq \mathbb{P}(A_n/n \in [a,a+da])$ is used for both continuous and discrete random variables throughout the paper. The dominant exponential behaviour of the probability measure corresponds to $\mathbb{P} \left( A_n / n = a \right) = e^{- n I(a) + o(n)}$ for large $n$. Usually, in large deviation applications, this is written as:
\be
\mathbb{P} \left( \frac{A_n}{n} = a \right) \asymp e^{-n I(a)} ,
\ee
where $\asymp$ represents asymptotic identity at logarithmic scale.
The rate function $I$ is continuous\footnote{In general $I$ is lower semi-continuous. This is equivalent to saying that it has closed level sets $\left\lbrace a: I(a) \leq c \right\rbrace$. See \cite{dembo2010large,den2008large} for details.} and its zeros represent typical values attained by $A_n/n$ in the thermodynamic limit $n \ra \infty$, while its tails characterize how likely fluctuations are to appear. 

One straightforwardly has LDPs for the time-additive observables $N_n/n$ and $J_n/n$ such that we can write the large deviation forms $P(\eta) \asymp e^{-n I(\eta)}$, and $P(j) \asymp e^{-n I(j)}$. The traditional way to prove an LDP for a general observable $A_n$ is to apply the G\"{a}rtner-Ellis theorem, which makes use of a Legendre-Fenchel transform in order to calculate the rate function
\be
\label{eq:lftransf}
I(a) = \sup_{s} \left( s a - \lambda(a) \right) ,
\ee
from the scaled cumulant generating function (SCGF) defined as
\be
\lambda(s) = \lim_{n \ra \infty} \frac{1}{n} \ln \mathbb{E}\left[e^{s A_n} \right] .
\ee
Note that this theorem requires that the SCGF exists and is differentiable.\footnote{Under such conditions, since the SCGF $\lambda$ is convex \cite{dembo2010large,touchette2009large}, the G\"{a}rtner-Ellis theorem ensures that the rate function $I$ is a good rate function. This means that it has compact level sets and the probability measure is exponentially tight. See \cite{dembo2010large} for details.} 

For the models $M_1$ and $M_2$, introduced in Sect. \ref{subsec:simplmod}, the SCGFs associated to $N_n/n$ and $J_n/n$ can be calculated straightforwardly. For the on-off process they are 
\be
\lambda_{M_1}(l) = \lambda_{M_2}(l) = \ln \left( r e^l + (1-r) \right) ,
\ee 
whereas for the current process we have 
\be
\lambda_{M_1}(k) = k^2 + \mu k ,
\ee and 
\be
\lambda_{M_2}(k) = \ln \left( r + (1-r)e^{k^2 + \mu k} \right) .
\ee 
Notice that $\lambda_{M_1}(k)$ is the SCGF associated to a random walk with no resets, and it often appears in the text. Throughout the manuscript we consistently use $l$ for the conjugate variables to $N_n/n$ and $k$ for the conjugate variable to $J_n/n$ to indicate implicitly the corresponding random variable without complicating the notation. [A similar convention applies for the arguments of rate functions.] All the functions introduced above are differentiable in the interior of their domains, thus in principle one can calculate the corresponding rate functions via the G\"{a}rtner-Ellis (\ref{eq:lftransf}). Analytically we can show for $M_1$ that $I_{M_1}(\eta) = \eta \ln \eta + (1 - \eta) \ln \left(1 - \eta \right) - \eta \ln r - \left( 1 - \eta \right) \ln \left( 1 - r \right)$, and $I_{M_1}(j) = \left( j - \mu \right)^2/4 $, whereas for $M_2$, although we have $I_{M_2}(\eta) = I_{M_1}(\eta)$, $I_{M_2}(j)$ can only be calculated numerically.

Making use of the G\"{a}rtner-Ellis theorem once again, it is also possible to show that an LDP holds for the joint probability density function $P(\eta,j)$. In order to do so, we need to find the SCGF $\lambda(l,k)$. For the model $M_1$, since $\mathbf{Y}_n$ and $\mathbf{X}_n$ are independent processes 
\be
\lambda_{M_1}(l,k) = \lambda_{M_1}(l) + \lambda_{M_1}(k) = \ln \left( r e^l + (1-r) \right) + k^2 + \mu k .
\ee 
However, for $M_2$ more care is needed. We calculate the moment generating function $G_{M_2}(l,k,n)$ directly using the definition of $G$ and conditioning the process $\mathbf{Y}_n$ on $\mathbf{X}_n$, 
\be
\label{eq:momgenjoint}
\begin{split}
G_{M_2}(l,k,n) &= \mathbb{E} \left[ e^{l N_n  + k J_n} \right] \\
&= \sum_{\mathbf{x}_n} \int_{\mathbf{y}_n \in \mathbb{R}^n} d \mathbf{y}_n \bigg( \mathbb{P}(\mathbf{Y}_n= (y_1,...,y_n)|\mathbf{X}_n=(x_1,...,x_n)) \\
&\;\;\;\;\; \times \mathbb{P}(\mathbf{X}_n= (x_1,...,x_n)) e^{l \sum_{i} x_i + k \sum_{i} y_i} \bigg) .
\end{split} 
\ee

First we exploit the independence in both processes and then the fact that the $X_i$s are identically distributed:
\be
\begin{split}
G_{M_2}(l,k,n) &= \int_{\mathbf{y}_n \in \mathbb{R}^n} d \mathbf{y}_n \sum_{x_i \in \left\lbrace 0,1 \right\rbrace} \mathbb{P}(X_i=x_i)e^{l x_i} \mathbb{P}(Y_i=y_i|X_i=x_i)e^{k y_i} \\
&= \int_{\mathbf{y}_n \in \mathbb{R}^n} d \mathbf{y}_n \left( re^{l} \delta (y_i) e^{k y_i} + (1-r) \mathbb{P}(Y_i=y_i|X_i=0)e^{k y_i} \right) \\
&= \prod_{i=1}^n \left( re^{l} + (1-r) \int_{y_i \in \mathbb{R}} dy_i \; P(y_i|X_i=0)e^{k y_i} \right)  \\
&= \left( re^{l} + (1-r) e^{\lambda_{M_1}(k)} \right)^n .
\end{split}
\ee
The rescaled limit of the logarithmic moment generating function is
\be
\lambda_{M_2}(l,k) = \lim_{n \ra \infty} \frac{1}{n} \ln G_{M_2}(l,k,n) = \ln \left( r e^l + (1-r)e^{\lambda_{M_1}(k)} \right) .
\ee
Hence, both $\lambda_{M_1}(l,k)$ and $\lambda_{M_2}(l,k)$ exist and are differentiable in the interior of their domains $D \times \mathbb{R}$. This is sufficient to state that LDPs hold for the joint probability density functions $P_{M_1}(\eta,j)$ and $P_{M_2}(\eta,j)$ with the associated rate functions obtained through Legendre-Fenchel transform
\be
\label{eq:LFtransfJoint}
I(\eta,j) = \sup_{l,k} \left( \eta l + j k - \lambda(l,k) \right) .
\ee
In fact, for $M_1$ it suffices to recall that $\mathbf{Y}_n$ and $\mathbf{X}_n$ are independent of each other, and this implies that $I_{M_1}(\eta,j) = I_{M_1}(\eta) + I_{M_1}(j)$. In contrast, for $M_2$ correlations between the top and the bottom layers do not allow us to proceed analytically and $I_{M_2}(\eta,j)$ can only be calculated numerically, either parametrically, exploiting Legendre duality $I_{M_2}(\lambda'(l),\lambda'(k)) = \lambda'(l) l + \lambda'(k) k - \lambda(l,k)$, or by direct implementation of the Legendre-Fenchel transform (\ref{eq:LFtransfJoint}).\footnote{Since $\lambda_{M_2}$ is differentiable these two methods give the same result.}

\subsection{Large deviation principle for the ratio}
\label{subsec:LDPbycontr}

We now turn to the main topic of the paper, showing that an LDP holds also for the extensive observable $n \Omega_n$ in the form
\be
\label{eq:ldpratio}
\lim_{n \ra \infty} \frac{1}{n} \ln \mathbb{P} \left( \Omega_n \in [\omega,\omega+d\omega] \right) = I(\omega).
\ee
This follows by the contraction principle \cite{touchette2009large,dembo2010large} from the LDP for the joint probability density function. 
The contraction principle is a powerful general technique which shows that an LDP is conserved under any continuous mapping, and here makes evident how the LDP for the ratio derives as a restriction on the bigger state space of $( N_n,J_n )$. More concretely the contraction principle emerges as a saddle-point approximation on the LDP for the probability density function $P(\eta,j)$. As a consequence an LDP holds also for the ratio with rate function
\be
\label{eq:LDPratio}
I(\omega) = \inf_{\substack{\eta \\ j = \omega \eta}} I(\eta,j) .
\ee
A caveat here is that, in fact, our mapping $\omega = j/\eta$ is continuous everywhere except at $\eta = 0$; this has important consequences to which we will return later.

\begin{figure}
    \centering
    \psfrag{a}{$\omega$}
    \psfrag{b}{$I$}
    \psfrag{c}{$\mu=-2$}
    \psfrag{d}{$\mu=-1$}
    \psfrag{e}{$\mu=0$}
    \psfrag{f}{$\mu=1$}
    \psfrag{g}{$\mu=2$}
    \subfloat[Model $M_1$]{{\includegraphics[width=5.5cm]{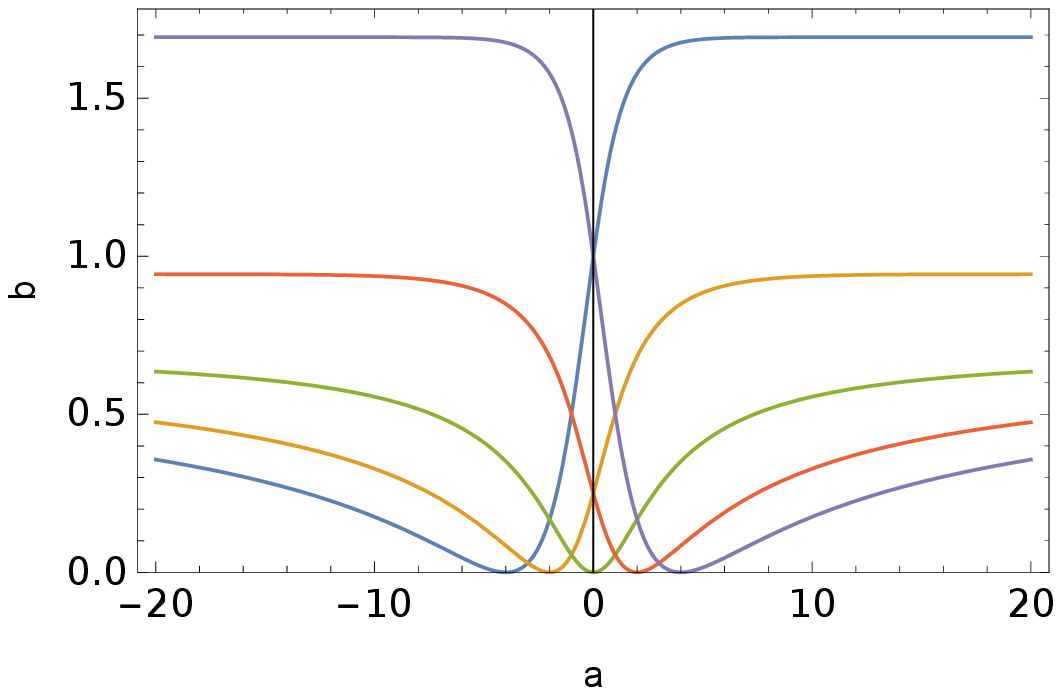} }}%
    \qquad
    \subfloat[Model $M_2$]{{\includegraphics[width=5.5cm]{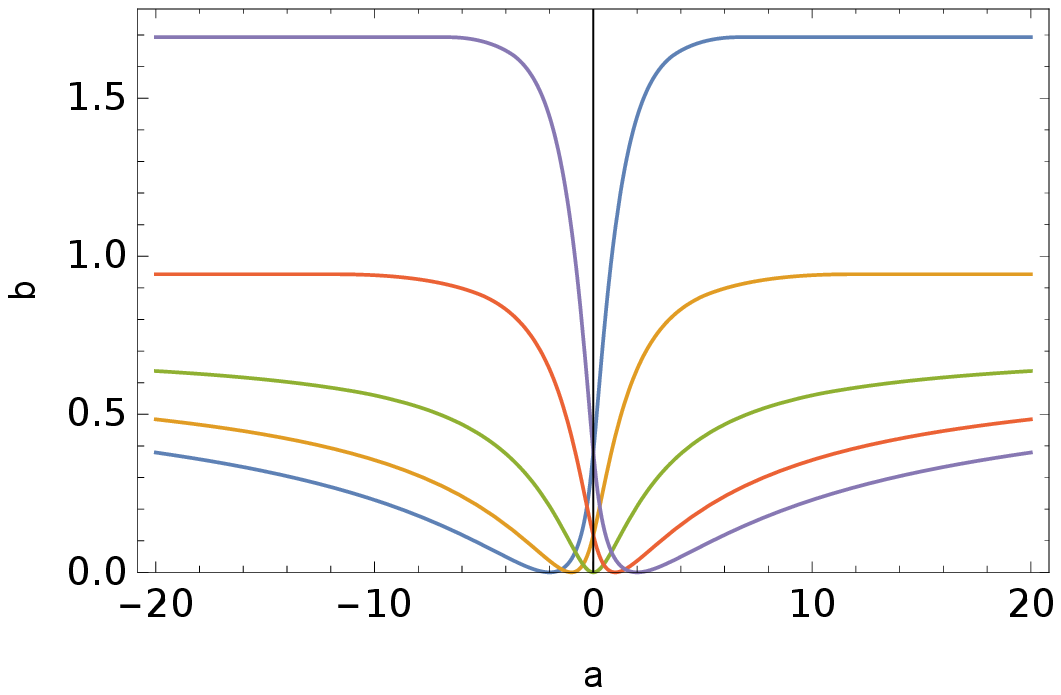} }}%
    \caption{Ratio rate function $I(\omega)$ calculated by contraction for (a) model $M_1$ and (b) model $M_2$. Curves for $\mu=-2$ (blue), $\mu=-1$ (orange), $\mu=0$ (green), $\mu=1$ (red), and $\mu=2$ (purple).}
    \label{fig:M1M2RatioRate}
\end{figure}

For the rate functions $I_{M_1}(\omega)$ and $I_{M_2}(\omega)$ the `infimum' in equation (\ref{eq:LDPratio}) involves transcendental equations, but we report in Fig. \ref{fig:M1M2RatioRate} the rate functions calculated numerically.
Note, as expected, that: (i) there is a unique zero representing the typical value taken by the ratio observable in the long-time (thermodynamic) limit $n \ra \infty$, which is easily calculated as $\hat{\omega}_{M_1} = \mu /r$ for the case of completely uncorrelated layers and $\hat{\omega}_{M_2} = (1-r) \mu /r$ for the correlated model, where the random walk in the top layer only hops on average for a fraction $1-r$ of steps; (ii) the fluctuations, represented by the non-zero values of the rate functions, are obviously not symmetric for $\mu \neq 0$; and (iii) the curves look smooth. [This last point is investigated in detail in Sect. \ref{sec:results}.] 

A more unusual feature of the ratio rate functions is the presence of horizontal asymptotes and the associated non-convexity.\footnote{This means that the rate function is a weak (not-good) rate function with level sets that are closed, but not compact \cite{dembo2010large,den2008large}. Such a weak rate function is a necessary and sufficient condition for the probability density function not being exponentially tight.} In fact, the horizontal asymptotes correspond to the limit $\eta \ra 0^+$, where the mapping is not continuous. 
Here rare events are realised by heavy tails which mask the linear exponential scaling of the ratio observable. Furthermore, we can straightforwardly obtain the position of the asymptotes through the argument that large deviations are realised in the least unlikely amongst all the unlikely ways \cite{den2008large}. In the case $\mu > 0$ such analysis shows that values of the rate function for $\omega \ra  \infty$ are given by the probability of a typical current $J_n = \hat{j}$ and $N_n \ra 0^+$, rather than $J_n \ra \infty$ and $N_n$ finite and non-zero. Similarly, in the case we look at $\omega \ra - \infty$, the rate function is given by the probability of $J_n \ra 0^-$ and $N_n \ra 0^+$, rather than requiring $J_n \ra -\infty$. Hence, the asymptotes read (in the case $\mu > 0$)
\be
\begin{split}
I(\omega \ra + \infty) &= I(\eta \ra 0^+) ,\\
I(\omega \ra - \infty) &= I(\eta \ra 0^+) + I(j \ra 0^-) .
\end{split}
\ee
Evidently $I(\omega \ra - \infty) > I(\omega \ra + \infty)$, corresponding to asymmetric fluctuations for $\mu > 0$. A reflected argument holds for $\mu < 0$, whereas for $\mu = 0$, due to the symmetry of the random walk, asymptotes are equivalent and fluctuations are symmetric. Note also that the non-convexity of the rate function is a feature that could not have been obtained by means of the G\"{a}rtner-Ellis theorem and Legendre-Fenchel transformation.

In closing this subsection, we remark that similar features have been observed in other ratio observables in the field of stochastic thermodynamics. In the work of Verley et al. \cite{verley2014unlikely}, and following papers \cite{verley2014universal,gingrich2014efficiency,proesmans2015stochastic,polettini2015efficiency}, the object of study is the ratio between the output work produced and the input heat absorbed in different systems representing nano-machines operating in the presence of highly fluctuating energy fluxes. In that case too the tails of the ratio rate function were found to tend to an asymptotic value. Indeed, we can understand this asymptotic behaviour in the rate function as a universal property of ratio observables when the denominator can, in principle, approach $0$. For instance, one can simply think of the ratio of two arbitrary Gaussian distributions \cite{marsaglia1965ratios,hinkley1969ratio,hinkley1970correction,marsaglia2006ratios}, 
which can be proven to be always composed of a mixture of two terms: a Cauchy unimodal distribution, and a bimodal distribution with heavy tails. Note that in this example, and the thermodynamic efficiency associated with nano-engines, the denominator in the ratio can have both positive \emph{and negative} fluctuations which generically lead to the presence of a \emph{maximum} in the rate function. In particular, this maximum marks a transition between a phase where fluctuations are generated by atypical values in the numerator, and a phase where they are generated by atypical values in the denominator. In contrast, in our case, the denominator can have only positive fluctuations so no such maximum appears.

\section{Results}
\label{sec:results}

In this section we apply the previously-introduced methods to analyse how the ratio observable behaves in some stochastic reset models. 

\subsection{Robustness and differentiability of ratio rate functions}
\label{subsec:RobRate}

So far we have seen that in toy models, e.g. $M_1$ and $M_2$, the ratio rate function appears smooth (everywhere differentiable). We are interested in understanding whether this is always the case even for genuine reset models with correlations. 

It is known that, at equilibrium, non-differentiable points in rate functions are connected to phase transitions. For instance, 
the microcanonical entropy of non-additive systems with long-range interactions, under mean-field-like approximations, can present non-differentiable points \cite{chavanis2002phase,bouchet2005classification,gross2004microcanonical,hovhannisyan2017complete}, signalling microcanonical first-order transitions. Furthermore, the appearance of cusps in SCGFs or in large deviation rate functions is also an important topic in nonequilibrium physics and can be understood as the appearance of dynamical phase transitions in the fluctuations. As one recent example, a non-differentiable point in the Freidlin-Wentzell (low-noise limit) large deviation rate function for the current of a driven diffusing particle on a ring has been identified \cite{nyawo2016large,proesmans2019large,mehl2008large}. In a similar spirit, in the next sub-sections, we will examine the smoothness of the ratio rate function for stochastic reset processes with dynamical phase transitions in the fluctuations of observables $J_n$ and $N_n$. 

In order to check if a ratio rate function $I(\omega)$ is differentiable we seek to find necessary and sufficient conditions for the appearance of non-differentiable points.
So far, many results are known regarding the regularity of functions coming from a variational calculation such as that of equation (\ref{eq:LDPratio}). In \cite{danskin1966theory}, and later in \cite{hogan1973directional,danskin2012theory}, sufficient conditions have been found such that $I(\omega) = \inf_{\eta} \tilde{I}(\eta,\omega)$ is $C^1(\mathbb{R})$, where $ \tilde{I}(\eta,\omega) \coloneqq I(\eta,j = \omega \eta)$ is $C^1(O)$, with $O \triangleq D \setminus \left\lbrace 0 \right\rbrace \times \mathbb{R}$.\footnote{Note that in the original formulation of \cite{danskin1966theory} $\eta$ must be in a closed bounded Euclidean space. To satisfy this condition we can always consider compact subsets of $D \setminus \left\lbrace 0 \right\rbrace$.}
The conditions to meet are: first, $\eta$ has to be minimized over a set that does not depend on $\omega$, and second, related to the implicit function theorem, the minimizer function $\eta(\omega)$, satisfying equation (\ref{eq:LDPratio}), has to be continuous and bijective. In our case, the first condition always holds so negation of the second one, meaning that the solution set of minimizers $\eta$ (for a particular $\omega$) is not singleton, is necessary for the function $I(\omega)$ to present jumps in its first derivative. We believe that for well-behaved $I(\eta,j)$ this necessary condition becomes also sufficient; one exception is the case where $I(\eta,0)$ itself has a non-singleton set of minima. 

In practice, this means that determining whether $I(\omega)$ is differentiable boils down to analysing
\be 
\frac{\partial \tilde{I}(\eta,\omega)}{\partial \eta} = 0 .
\ee
If for a particular $\omega^*$ this equation is verified for more than a single value of $\eta \in \left( 0,1 \right]$, then a non-differentiable point should appear at $I(\omega^*)$. We will see in Appendix \ref{app:Mb1} an example where this can be checked analytically. However, often an analytical form of $\tilde{I}(\eta,\omega)$ is not available and in such cases we conduct a numerical analysis, discretizing the domain of $\omega$ and plotting the locus of minimizing points $( \eta, \eta \omega )$ of equation (\ref{eq:LDPratio}) on the joint rate function $I(\eta,j)$. In general, if the set of minimizers $\eta$ is not always singleton, the locus $( \eta, \eta \omega )$ on $I(\eta,j)$ presents a linear section of values $\eta$ satisfying the minimization condition, and therefore $I(\omega) \notin C^1(\mathbb{R})$. Such a feature is seen in the analytical example of Appendix \ref{app:Mb1} and is related to the appearance of a `mode-switching' dynamical phase transition in the generation of fluctuations; it is useful to compare Fig. \ref{subfigb:b1ZerosRatioRate} showing the phase transition with Figs. \ref{subfigb:aZerosRatioRate} and \ref{subfigb:bZerosRatioRate} of the main text where no such transition is present.

As well as smoothness of the ratio rate function we are interested in whether it retains its general shape under the addition of interactions in genuine reset processes. Loosely speaking, we will use the term `robust' for the case where the qualitative shape retains the salient features discussed in Sect. \ref{subsec:LDPbycontr}. This is reminiscent of the concept of `universality' in other areas of statistical physics. 

\subsection{Finite-time correlations}
\label{subsec:Moda}

Our analysis begins with the model of \cite{harris2017phase}. In this model, the combination of reset and finite-time correlations in the random walk generates dynamical phase transitions (DPTs) in the observable $J_n$. These are distinguished by the analyticity of the SCGF: for first-order DPTs the SCGF is not differentiable, whereas for continuous DPTs it is. Here, DPTs are interpreted as transitions between fluctuations that involve resets and fluctuations that do not. The Legendre-Fenchel transform can be applied to the differentiable branches of the SCGF, and, as a consequence, the rate function so obtained will present a gap for the set of values for which the derivative of the SCGF does not exist. It is customary in statistical mechanics to extend the region over which the Legendre-Fenchel transform is defined by a Maxwell construction, i.e. by drawing a linear section connecting the two branches of the rate function. In general, the function so derived is the convex-hull of the true rate function, but for finite-time correlations, because of subadditivity, it is known to be exactly the true rate function. In \cite{harris2017phase} the linear sections appearing in the current rate functions correspond to mixed regimes where typical trajectories switch between periods with resets and periods with no resets. In the following, we want to understand how these DPTs influence the ratio observable $\Omega_n$.

\subsubsection{Model $M_a$}

As in $M_1$ and $M_2$ the bottom layer is a discrete-time stochastic process $\mathbf{X}_n = \left( X_1,X_2,...,X_n \right)$ composed of $n$ Bernoulli random variables of parameter $r$, and the top layer is a discrete-time (but continuous-space) random walk $\mathbf{Y}_n = \left( Y_1,Y_2,...,Y_n \right)$. At time step $l$ after a reset, the random walk takes a jump according to an $l$-dependent Gaussian density function with mean $\mu$ and variance $\sigma^2 = 2(1-B/(l+d))$, a function of parameters $d$ and $0 < B \leq d+1$. If time goes on, and reset events do not happen, the variance of the Gaussian distribution increases monotonically towards the asymptote $\sigma^2 = 2$ as $l \ra \infty$. In this model we focus on the long-time behaviour of the observables introduced in Sect. \ref{subsec:simplmod}: $N_n$, $J_n$, and the ratio $\Omega_n$. As $J_n$ is correlated with $N_n$, this leads naturally to short range correlations, and although DPTs are not present in $N_n$, they are in $J_n$. 

\subsubsection{Joint scaled cumulant generating function}
\label{subsec:JointMa}

In order to derive the function $I_{M_a}(\omega)$, characterising the exponential scaling of $\Omega_n$, we first need to derive the joint rate function $I_{M_a}(\eta,j)$, which can be done by applying the standard procedure of Sect. \ref{sec:methods} for LDPs of empirical means. It is difficult to calculate the SCGF $\lambda_{M_a}(l,k)$ directly from the definition $\lim_{n \ra \infty} (1/n) \ln \mathbb{E} \left[ e^{l N_n + k J_n} \right]$, as there is no factorization property. 
For this reason we rely on a different technique originally applied to study the free-energy of long linear chain molecules \cite{lifson1964partition}, such as DNA \cite{poland1966occurrence}, and more recently used in the context of continuous-time \cite{meylahn2015large} and discrete-time reset processes \cite{harris2017phase}. The strategy is to rewrite the moment generating function $G_{M_a}^r(l,k,n)$ as a convolution of independent `microscopic' contributions and then to take its discrete Laplace transform (or $z$-transform) $\tilde{G}_{M_a}^r(l,k,z)$. This method is tantamount to working in the grand-canonical ensemble in time, where $z$ represents the fugacity, and allows us to relax the constraint we would have on summing over paths of a certain length when calculating the moment generating function directly. The SCGF $\lambda_{M_a}(l,k)$ is then obtained as the natural logarithm of the radius of convergence $z^*(l,k)$ of $\tilde{G}_{M_a}^r(l,k,z)$.  

In our set-up the microscopic moment generating function for a sequence of $n-1$ non-reset steps (along with $n-1$ random walk jumps) followed by one reset step is \footnote{One can also define two different microscopic generating functions for sequences characterised by only non-reset steps and only reset steps. As in \cite{harris2017phase}, this approach is particularly useful when the probability of reset does not depend on the time elapsed since last reset. However, our generating function (\ref{eq:GenFunSeq}), built on the more general framework of renewal theory, allows the consideration of different scenarios in the on-off process \cite{harris2019thermodynamic}.}
\be
\label{eq:GenFunSeq}
W(l,k,n) = \mathbb{E} \left[ e^{l N_n + k J_n} \right] = r e^l (1-r)^{n-1} e^{(n-1) \left( k^2 + \mu k \right) - B k^2 \left( H_{n-1+d} - H_d \right) } ,
\ee
where $n \geq 1$ and $H_n = \sum_{k=1}^n 1/k$ is the truncated harmonic series. Note that we exclude microscopic sequences of zero length by enforcing $W(l,k,0) = 0$. The convolution of the microscopic moment generating functions returns the generating function of the whole process. Notice that this procedure assumes that the process always finishes with a reset step; this assumption is expected to make no difference in the infinite-time limit, at least in the case of finite moments in the distribution of inter-reset times.

We can now calculate the $z$-transform $\tilde{G}_{M_a}^r(l,k,z)$. First we distribute the $n$ factors of $z^{-n}$ among the microscopic sequences and then we change the order of summation over $n$ and $s$ as follows:

\be
\label{eq:zTransfWholeProc}
\begin{split}
\tilde{G}_{M_a}^r(l,k,z) &= \sum_{n=1}^{\infty} z^{-n} \sum_{s=1}^{n} \sum_{\left\lbrace i_{\sigma} \right\rbrace} \prod_{\sigma=1}^s W(l,k,i_{\sigma}) \\
&= \sum_{s=1}^{\infty} \sum_{n=s}^{\infty} \sum_{\left\lbrace i_{\sigma} \right\rbrace} \prod_{\sigma=1}^s W(l,k,i_{\sigma}) z^{-i_{\sigma}} .
\end{split}
\ee
The $\sum_{\left\lbrace i_{\sigma} \right\rbrace}$ is interpreted as a sum over all possible configurations of $s$ sequences of length $\left\lbrace i_1, i_2, ..., i_s \right\rbrace$ with the constraint that $\sum_{\sigma} i_{\sigma} = n$. This restriction can be dropped when first summing over all possible values of $s$, allowing us to carry out each sum over $i_{\sigma}$ independently
\be
\begin{split}
\tilde{G}_{M_a}^r(l,k,z) &= \sum_{s=1}^{\infty} \prod_{\sigma=1}^s \sum_{i_{\sigma}=1}^{\infty} W(l,k,i_{\sigma}) z^{-i_{\sigma}} \\
&= \frac{\tilde{W}(l,k,z)}{1-\tilde{W}(l,k,z)} .
\end{split}
\ee
In the last step a geometric series appears and $\tilde{W}(l,k,z) \coloneqq \sum_{n=1}^{\infty} W(l,k,n) z^{-n}$ denotes the $z$-transform of $W(l,k,n)$. Notice again that our choice of indexing excludes zero-length trajectories; this affects only the `boundary' term in the numerator. We remark that $\tilde{W}(l,k,z)$ converges only if $z \geq z_c(k) \coloneqq (1-r)e^{k^2+\mu k}$, where $\ln z_c$ corresponds to the SCGF of an i.i.d.\ random walk in the non-reset limit.

As mentioned, the SCGF $\lambda_{M_a}(l,k)$ corresponds to the natural logarithm of the largest real value $z^*(l,k)$ at which $\tilde{G}_{M_a}^r(l,k,z)$ diverges. This can be identified with $\hat{z}$, the largest real solution of the equation $\tilde{W}(l,k,z) = 1$, when $\hat{z} \geq z_c$; otherwise we have directly $z^* = z_c$. The change from convergent to divergent $\tilde{W}(l,k,z)$ corresponds to a phase transition in the reset process, between a regime where current fluctuations are optimally realised in the presence of reset events, and a regime where current fluctuations are realised by trajectories with no reset events at all \cite{harris2017phase,richard2004poland}. 
As discussed in the previous section, if the obtained SCGF $\lambda_{M_a}(l,k)$ is differentiable, the joint large deviation rate function $I_{M_a}(\eta,j)$ can be derived by Legendre-Fenchel transform (\ref{eq:LFtransfJoint}). \footnote{It may turn out that $\lambda_{M_a}(l,k)$ has non-differentiable points which mark first-order phase transitions; in this case the Legendre duality does not hold everywhere in the domain $D \times \mathbb{R}$ \cite{touchette2005legendre} restricting the validity of LDPs to differentiable regions.}

\subsubsection{Ratio observable}
\label{subsubsec:BRatioContr}

\begin{figure}
    \centering
    \psfrag{a}{$\omega$}
    \psfrag{b}{$I$}
    \psfrag{c}{$B=0.5$}
    \psfrag{d}{$B=1.0$}
    \psfrag{e}{$B=2.5$}
    \psfrag{f}{$B=4.0$}
    \subfloat[Ratio rate functions $I_{M_a}(\omega)$]{{\includegraphics[width=5.5cm]{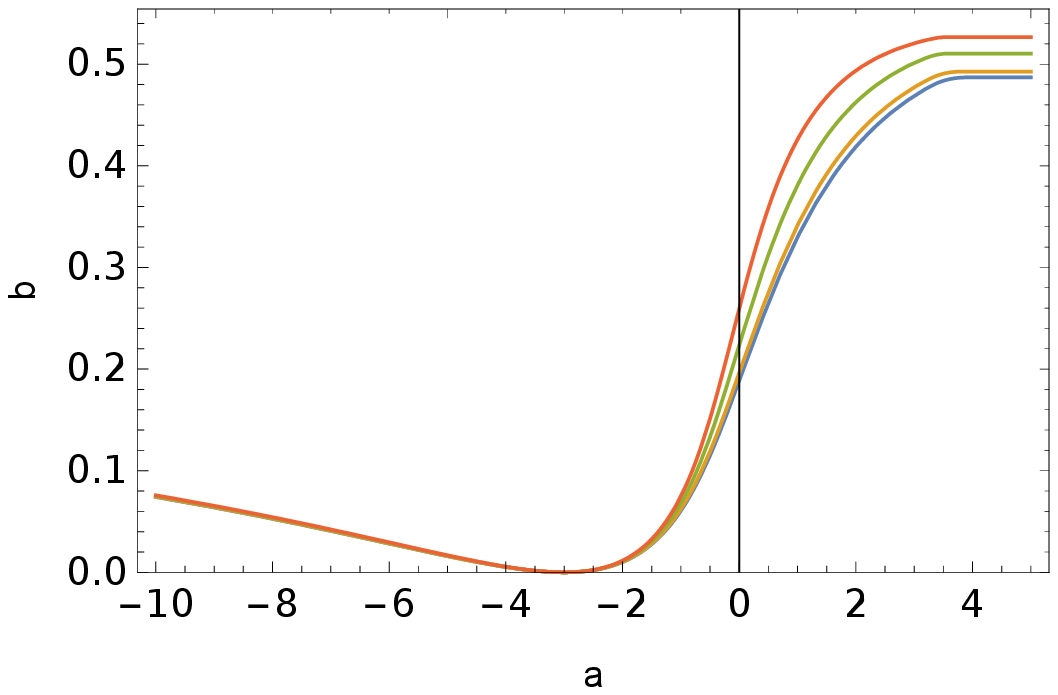}}\label{subfiga:aRatioRateAsymm}}%
    \qquad
    \subfloat[Locus of minimizers from contraction]{{\includegraphics[width=5.5cm]{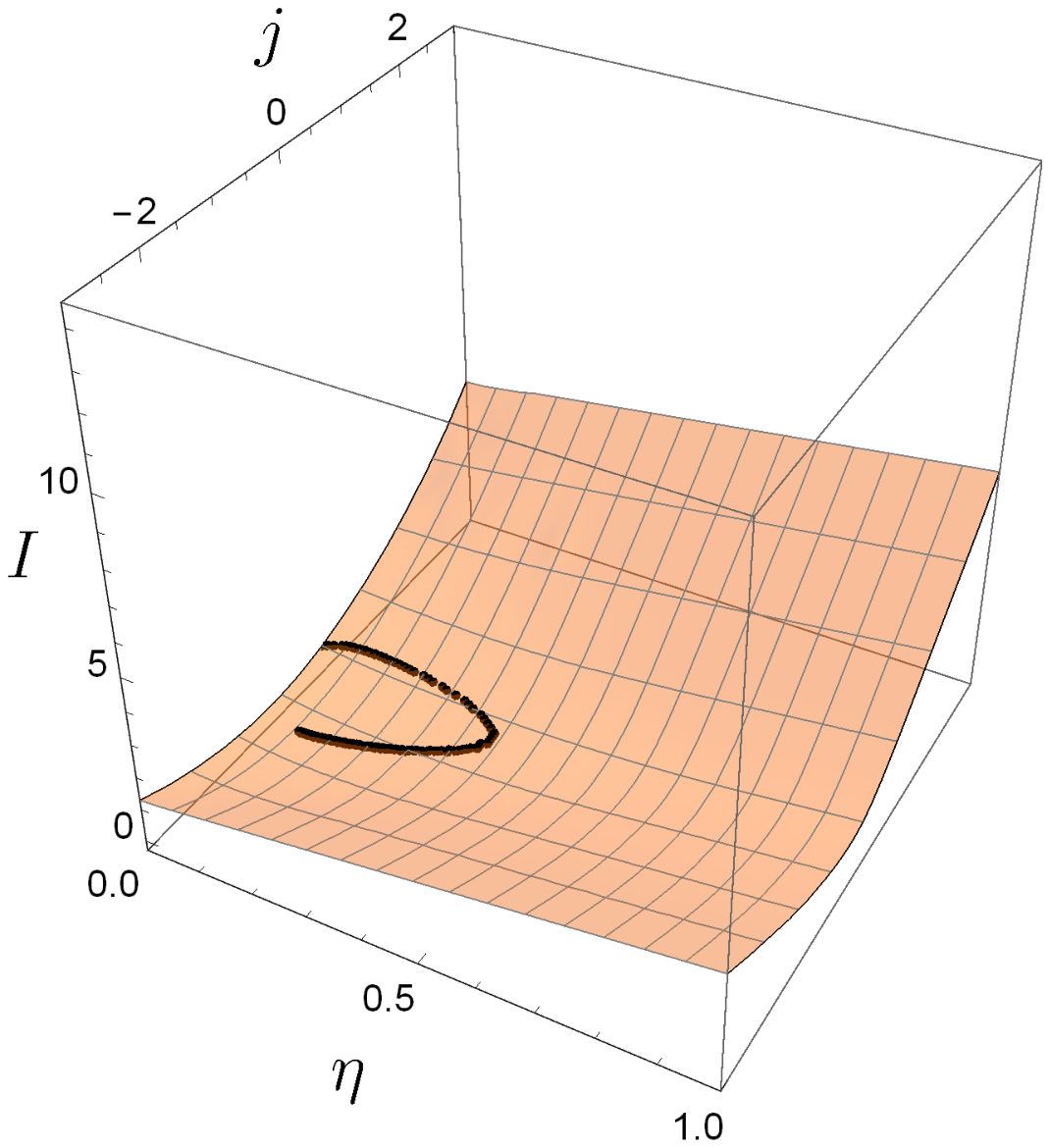} }\label{subfigb:aZerosRatioRate}}%
    \caption{(a) Ratio rate function $I_{M_a}(\omega)$ for $B= 0.5,1.0,2.5,4.0 $ (from bottom to top). (b) Locus of minimizers $( \eta, \omega \eta )$ satisfying the contraction principle (\ref{eq:LDPratio}) for $\omega \in [-20,20]$ and $B=2.5$ depicted on the surface of the joint rate function $I(\eta,j)$. Numerics for $\mu=-1$, $r=0.25$, and $d=10$.}
    \label{fig:afinal}
\end{figure}

By the contraction principle (\ref{eq:LDPratio}) on $I_{M_a}(\eta,j)$ we derive $I_{M_a}(\omega)$, focusing on the asymmetric case $\mu \neq 0$.
In Fig. \ref{subfiga:aRatioRateAsymm} we plot the ratio rate functions for a random walk with $\mu = -1$, and varying $B$. Notice that the left tails all coincide, indeed such fluctuations are obtained taking $\eta \ra 0^+$, regardless of the current, which stays in its typical state $\hat{j} = (1-r)\mu$ for any $\omega \leq \hat{\omega}_{M_a} = (1-r)\mu/r$. For any value of the parameters $\mu$ and $B$, the rate functions $I_{M_a}(\omega)$ in Fig. \ref{subfiga:aRatioRateAsymm} look robust, i.e. the qualitative features are unchanged under the appearance of DPTs in the fluctuations of the current $J_n$, generated by finite-time correlations.

$I_{M_a}(\omega)$ also looks differentiable. As argued in Sect. \ref{subsec:RobRate}, this can be investigated looking at the locus of minimizers $( \eta, \eta \omega )$, satisfying equation (\ref{eq:LDPratio}), on the joint rate function. We note that only the presence of first-order DPTs in the process is translated into the appearance of linear regions in $I_{M_a}(\eta,j)$, and only these could in principle influence the fluctuations of the observable $\Omega_n$. On increasing the parameter $B$ these linear regions extend and get closer to the bottom of the joint rate function, where contraction minimizers $( \eta, \eta \omega )$ lie. However, this does not affect the variational calculation much. We report in Fig. \ref{subfigb:aZerosRatioRate} example minimizers $( \eta, \eta \omega )$ for the case $B=2.5$. It is evident that the locus of minimizing points is a curve which stays close to the minimum of $I_{M_a}(\eta,j)$ where linear sections from first-order DPTs extend only in pathological cases (e.g. $r \ra 1$, $d \ra \infty$ and $B = O(d)$).

In order to gain more general understanding about the robustness and differentiability of the ratio rate function when finite-time correlations generate DPTs, we also investigated a rather unphysical model (denoted by $M_{a1}$) characterised by having a first-order DPT in the on-off process in the bottom layer uncoupled from the random walk in the top layer. The joint SCGF in this case is artificially constructed and is $\lambda_{M_{a1}}(l,k) = \lambda_{M_{a1}}(l) + \lambda_{M_{a1}}(k)$, with
\be
\begin{split}
\lambda_{M_{a1}}(l) &=
\begin{cases}
-\frac{1}{16} & \text{if} \;\; l \leq -\frac{1}{4} \\
l^2 + \frac{l}{2} & \text{if} \;\; l \in [-\frac{1}{4},b] \\
l + b^2 - \frac{b}{2} & \text{if} \;\; l \geq b ,
\end{cases} \\
\lambda_{M_{a1}}(k) &= k^2 + \mu k .
\end{split}
\ee
Here $0 \leq b \leq 1/4$ is a parameter which allows us to move the first-order DPT. Calculating analytically the rate function $I_{M_{a1}}(\eta)$ we see that for small but finite $b$ the linear section extends close to the minimum without actually reaching it. We see that in this case the ratio rate function $I_{M_{a1}}(\omega)$ is robust and presents a unique typical state. The limiting case $b = 0$ is pathological in the sense that $I_{M_{a1}}(\eta)$ has a flat section at zero leading to a corresponding flat section in $I_{M_{a1}}(\omega)$. However, even here the ratio rate function is differentiable.

\subsubsection{Numerical checks}
\label{subsec:NumChecks}

In deriving the SCGF as the natural logarithm of the convergence radius $z^*(l,k)$, we assume that any non-analyticities in pre-factors in the moment generating function $G_{M_a}^r(l,k,n)$ do not affect its exponential behaviour in the limit $n \ra \infty$. \footnote{As shown in \cite{gupta2017stochastic}, when calculating a joint probability density function by means of a Bromwich integral, non-analyticities in the characteristic function can create a singularity contour that is crossed by the saddle-point path. In such a case one should consider both contributions (saddle and branch-cut) in the calculation of the Bromwich integral.} To show that such pre-factors do not play a role here, we make use of an inverse numerical $z$-transform of $\tilde{G}_{M_a}^r(l,k,z)$ to check that the directly calculated moment generating function $G_{M_a}^r(l,k,n)$ approaches $z^*(l,k)$ smoothly in the limit $n \ra \infty$.

The inverse $z$-transform of $\tilde{G}_{M_a}^r(l,k,z)$ is defined as 
\be
G_{M_a}^r(l,k,n) \coloneqq \frac{1}{2 \pi i} \oint_{\mathcal{C}} \tilde{G}_{M_a}^r(l,k,z) z^{n-1} dz .
\ee
However, numerical integration may lead to inaccurate results and hence we make use of two other techniques as explained in \cite{merrikh2014two}: the first method is algebraic, based on truncating the $z$-transform, the second method relies instead on a discrete Fourier transform. We refer to Appendix \ref{app:InvzTransf} for further details on the methods.

\begin{figure}
    \centering
    \psfrag{a}{$k$}
    \psfrag{b}{$\ln z^*$}
    \psfrag{c}{$\text{Inv. Fourier}$}
    \psfrag{d}{$\text{Alg}$}
    \subfloat[$\mu=-1$, $r=0.25$, $B=0.5$, $d=10$]{{\includegraphics[width=5.5cm]{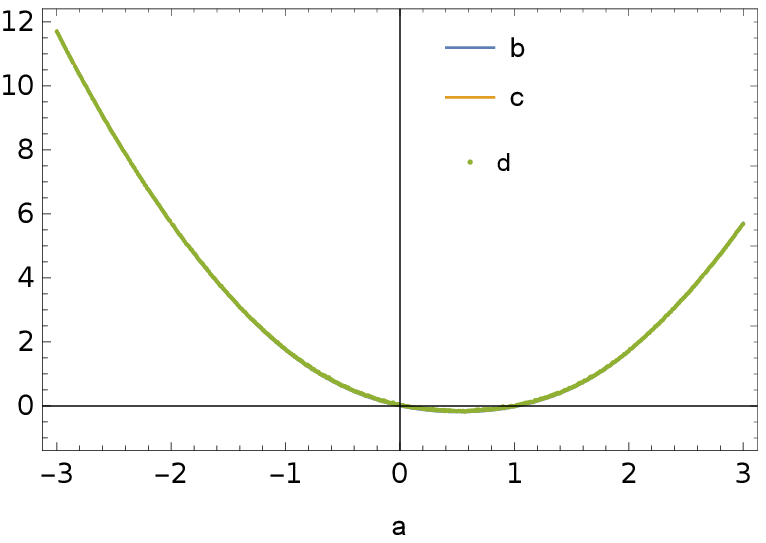} }}%
    \qquad
    \subfloat[$\mu=-1$, $r=0.25$, $B=2.5$, $d=10$]{{\includegraphics[width=5.5cm]{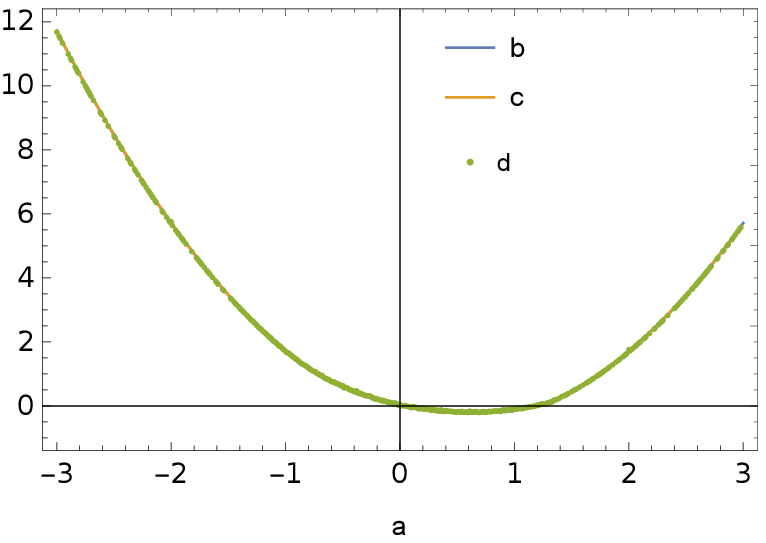} }}%
    \caption{Comparison of SCGFs obtained as $\ln z^*(l,k)$ and through inverse $z$-transforms using the algebraic method (Alg), and the inverse Fourier transform (Inv. Fourier); (a) differentiable SCGF, (b) SCGF non-differentiable at $k \simeq 1.25$. Curves are not distinguishable by eye.}
    \label{fig:aCompSCGFInvFour}
\end{figure}

In Fig. \ref{fig:aCompSCGFInvFour} we compare the SCGF $\lambda_{M_a}^r(l,k)$ calculated as the natural logarithm of the convergence radius of $\tilde{G}_{M_a}^r(l,k,z)$ with the rescaled natural logarithm of the approximated moment generating function $G_r(l,k,n)$ obtained using the methods explained above, for cases and with and without a first-order DPT. As computation becomes daunting quite fast, we report the comparison only for a subset of the domain of $\lambda_{M_a}(l,k)$.
In both cases there is a very good matching between curves, suggesting that pre-factors in the moment generating function $G_{M_a}^r(l,k,n)$ do not influence our study, and the obtained LDPs for the joint observable $( N_n,J_n )$ are correct.

\subsection{Infinite-time correlations}
\label{subsec:Modb}

In our models so far we have seen that short-range correlations, although they may generate DPTs in fluctuations of the current or number of resets, do not have much influence on the asymptotic fluctuations of the observable $\Omega_n$, whose rate function is robust and stays differentiable. Now we extend the analysis to long-range correlations. We present here a model where infinite-time correlations appear in the bottom layer, representing the on-off process, and extend to the coupled random walk in the top one. In Appendix \ref{app:Mb1} we consider a similar artificial model where we remove the coupling between the two layers, allowing us to carry out some analytical calculations and make an illuminating comparison.

\subsubsection{Model $M_b$}

Differently from the models $M_1$ and $M_2$ of Sect. \ref{subsec:simplmod} and model $M_a$ of Sect. \ref{subsec:Moda}, here the bottom layer is composed of two discrete-time stochastic processes glued together: $n$ Bernoulli random variables of parameter $r$, $\mathbf{X}_n=\left( X_1,X_2,...,X_n \right)$, and a `stiff' block of another $n$ variables $\mathbf{X}_n'=\left( X_{n+1},X_{n+2},...,X_{2n} \right)$ either all $0$, or all $1$ with equal probability. Note that both the blocks are extensive in time. In the top layer a discrete-time and continuous-space random walk composed of $\mathbf{Y}_n = \left( Y_1,Y_2,...,Y_n\right)$ followed by $\mathbf{Y}_n' = \left( Y_{n+1},...,Y_{2n} \right)$ is coupled with the bottom process. If $X_i=0$ the random walk takes a jump of non-zero length according to a Gaussian density function of mean $\mu$ and variance $\sigma^2 = 2$; on the other hand when reset occurs $X_i=1$ and $Y_i = 0$. Besides the label $M_b$, we propose the name of two-block reset model to refer to this stochastic reset process. Indeed, the bottom on-off process is similar in spirit to the so-called two-block spin model introduced in \cite{touchette2008simple}; the first block of steps $\mathbf{X}_n$ plays the role of a classical paramagnet, whereas the second half $\mathbf{X}_n'$ is analogous to a ferromagnet and brings infinite-time correlations both in the bottom layer and, as a consequence of the coupling, in the top one. 

\begin{figure}
    \centering
    \subfloat[Reset rate function]{
	\psfrag{a}{$\eta$}
    \psfrag{c}{$I_{M_b}^{1/2}$}
    \psfrag{d}{$\text{Conv}(I_{M_b}^{1/2})$}
    {\includegraphics[width=5.4cm]{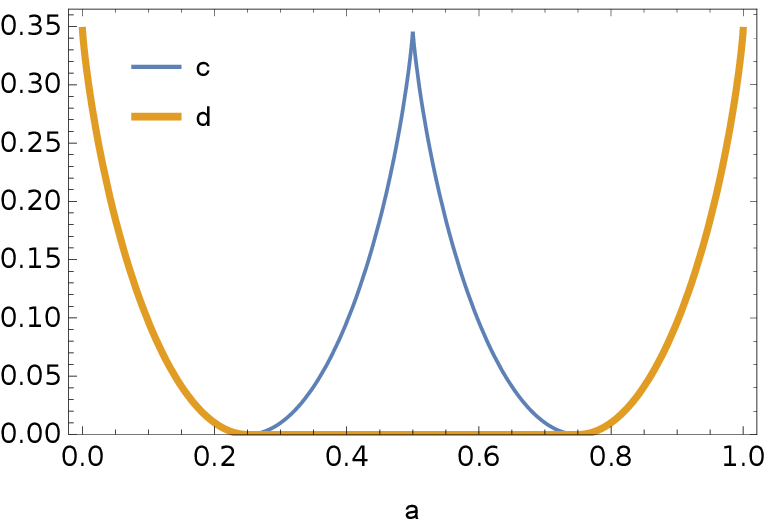}}\label{subfiga:bResRate}}%
    \qquad
    \subfloat[Current rate functions $I_{M_b}(j)$]{
	\psfrag{a}{$j$}
    \psfrag{b}{$I$}
    \psfrag{c}{$\mu = -2$}
    \psfrag{d}{$\mu = 0$}        
    {\includegraphics[width=5.4cm]{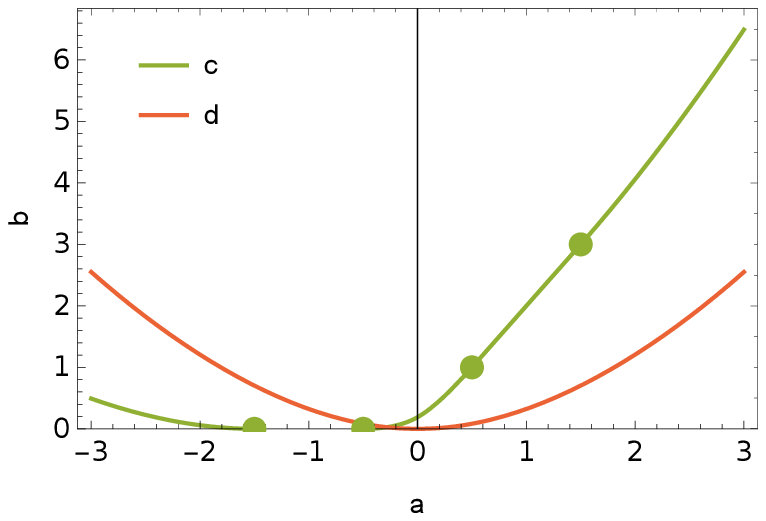} }\label{subfigb:bCurrRate}}%
    \caption{(a) Rate function $I_{M_b}^{1/2}(\eta)$ for the empirical mean number of reset steps plotted with its convex envelope $\text{Conv}(I_{M_b}^{1/2}(\eta))$. (b) Rate functions $I_{M_b}(j)$ for the empirical mean current  plotted for different values of $\mu$. DPTs are marked with straight lines delimited by coloured dots. Plots obtained for $r=0.5$.}%
    \label{fig:bRate}%
\end{figure}

If the Bernoulli parameter is $r=1/2$, our model is directly equivalent to the two-block spin model in \cite{touchette2008simple}, where reset steps correspond to up spins and non-reset steps correspond to down spins. In this case, we can obtain the large deviation rate function $I_{M_b}^{1/2}(\eta) = -\lim_{n \ra \infty} 1/(2n) \mathbb{P}(N_{2n}/(2n) = \eta)$ by following the derivation in \cite{touchette2008simple}. Specifically, we map the energy per spin $u$ into the mean number of reset steps $\eta$ according to $\eta = 1+u/2$ and, as in \cite{touchette2009large}, reflect and translate the microcanonical entropy $s(u)$ by $(1/2) \ln |\Lambda|$, where $|\Lambda| = 2$ is the state-space cardinality of the Bernoulli random variables. This leads to
\be
\label{eq:rateModBreset}
\begin{split}
I_{M_b}^{1/2}(\eta) &= \frac{\ln 2}{2}  - s(\eta) \\
&=
\begin{cases}
	  \frac{\ln 2}{2} & \eta = 0 \\
      \frac{\ln 2}{2} - \frac{2\eta-1}{2} \ln \left( 1-2\eta \right) + \eta \ln 2\eta & \eta \in (0,\frac{1}{2}] \\
      \frac{\ln 2}{2} - (\eta-1) \ln \left( 2-2\eta \right) + \frac{2\eta - 1}{2} \ln \left( 2\eta - 1 \right) & \eta \in (\frac{1}{2},1] ,\\
   \end{cases} 
\end{split} 
\ee
which we plot in Fig. \ref{subfiga:bResRate} along with its convex envelope. Notice that $I_{M_b}^{1/2}(\eta)$ has two minima $\hat{\eta}_1 = r/2$ and $\hat{\eta}_2 = (1+r)/2$, corresponding to the boundaries of the flat region of zeros in its convex envelope. 
Although in the general case $r \neq 1/2$, the microcanonical description breaks down, as microstates are no longer equally likely, it is still possible to calculate the rate function $I_{M_b}(\eta)$ from a probabilistic point of view. The derivation begins with conditioning $\mathbb{P}(N_{2n} = 2 n \eta)$ on the appearance of a block $\left( X_{n+1},X_{n+2},...,X_{2n} \right)$ of either all reset steps or all non-reset steps. This breaks the ergodicity, now either $\eta \in [0,1/2]$ or $\eta \in (1/2,1]$, and everything boils down to calculating the probability that in the first block $\left( X_1,X_2,...,X_n \right)$ there are either $n(2\eta -1)$ or $2n \eta$ reset steps. The number of reset steps follows a binomial distribution, thus making use of Stirling's approximation we get
\be
I_{M_b}(\eta) = 
\begin{cases}
	  -\frac{\ln (1-r)}{2}  & \eta = 0 \\
      \frac{1-2\eta}{2} \left[ \ln \left( 1-2\eta \right) - \ln (1-r) \right] + \eta \left[ \ln 2\eta - \ln r \right] & \eta \in (0,\frac{1}{2}] \\
      (1-\eta) \left[ \ln \left( 2-2\eta \right)  - \ln (1-r) \right] + \frac{2\eta - 1}{2} \left[ \ln \left( 2\eta - 1 \right) - \ln r \right] & \eta \in (\frac{1}{2},1] .
   \end{cases} 
\ee
Obviously, this recovers (\ref{eq:rateModBreset}) in the case $r=1/2$. Furthermore, as expected, $I_{M_b}(\eta)$ is a non-convex function, which is a consequence of long-range correlations in the model. Indeed, similarly to \cite{touchette2008simple}, adding an extensive block of steps which are either all reset or all non-reset, makes the model a `switch' between two different phases. Naturally, since the top layer is coupled with the bottom one, the phase transition appearing in the on-off process is inherited also by the random walk and this is reflected in the joint large deviation structure. From the joint SCGF $\lambda_{M_b}(l,k)$, calculated below, one obtains the current SCGF $\lambda_{M_b}(k) = \lim_{n \ra \infty} 1/(2n) \ln \mathbb{E}\left[ e^{2n k j} \right]$, which reads
\be
\lambda_{M_b}(k) = 
\begin{cases}       
      \frac{k^2 + \mu k}{2} + \frac{1}{2} \ln \left( r + (1-r)e^{k^2+\mu k} \right) & k^2 + \mu k > 0 \\
      \frac{1}{2} \ln \left( r + (1-r)e^{k^2+\mu k} \right) & k^2 + \mu k \leq 0 .
   \end{cases} 
\ee
Since $\lambda_{M_b}(k)$ has non-differentiable points, the Legendre-Fenchel transform only recovers the convex hull of the true current large deviation rate function. In fact, here the transform can only be done numerically; we show the result in Fig. \ref{subfigb:bCurrRate} for two different values of the parameter $\mu$ and $r = 1/2$. It is easy to prove that, provided $\mu \neq 0$, there are two jump discontinuities in the derivative of $\lambda_{M_b}(k)$. They arise at $k^*_1 = 0$, and at $k^*_2 = -\mu$, and are also evident in Fig. \ref{subfigb:bCurrRate} where $I_{M_b}(j)$ possesses linear sections with slope $k^*_1$ and $k^*_2$. In particular we note that the flat section of zeros is bounded by the typical values for the current $\hat{j}_1 = (1-r)\mu/2$ and $\hat{j}_2 = (2-r)\mu/2$.

Summing up, the main property of this model is the appearance of a DPT in the bottom layer, where long-range correlations break the ergodicity causing the system trajectories to be characterised by having either $\eta \in [0,1/2]$ or $\eta \in (1/2,1]$. In our reset process, since the random walk in the top layer is coupled with the on-off layer, the phase transition is inherited by the random walk, provided that $\mu \neq 0$. [If the random walk is symmetric, it will keep taking jumps of mean length $0$, and these do not bring any extensive contribution to the observable $J_{2n}$, regardless of the phase manifest in the bottom layer.] Below we consider how this behaviour is reflected in the observable $\Omega_{2n}$.

\subsubsection{Joint scaled cumulant generating function}

We first calculate the moment generating function $G_{M_b}^r(l,k,2n)$ by using its definition and introducing the auxiliary random variable $S \sim \text{Bernoulli}(1/2)$ characterising the nature of the block $\left( X_{n+1},X_{n+2},...,X_{2n} \right)$. This allows us to write the two observables of interest as $N_{2n} = \sum_{i=1}^n X_i + n S$ and $J_{2n} = \sum_{i=1}^n Y_i + (1-S) \sum_{i=1}^n Y_{n+i}$. The calculation follows by recognising that we can split the whole expectation value into two independent factors: one related to the process composed of Bernoulli random variables $\mathbf{X}_n$ in the bottom layer, and one related to the `stiff' bit $\mathbf{X}_n'$. This yields
\be
\begin{split}
G_{M_b}^r(l,k,2n) &= \mathbb{E} \left[ e^{2n (l N_{2n} + k J_{2n})} \right] \\
&=
\sum_{\mathbf{x}_n} \int_{\mathbf{y}_n \in \mathbb{R}^n} d \mathbf{y}_n \bigg( \mathbb{P}(\mathbf{Y}_n = \left( y_1,...,y_n \right)| \mathbf{X}_n = \left( x_1,...,x_n \right)) \\
&\;\;\;\;\;\times \mathbb{P}(\mathbf{X}_n = \left( x_1,...,x_n \right))  e^{l \sum_{i=1}^n x_i} e^{k \sum_{i=1}^n y_i} \bigg) \\
&\;\;\;\;\;\times \sum_{s \in \left\lbrace 0,1 \right\rbrace} \int_{\mathbf{y}_n' \in \mathbb{R}^n} d \mathbf{y}_n' \bigg( \mathbb{P}(\mathbf{Y}_n' = \left( y_{n+1},...,y_{2n} \right) | S=s) \\
&\;\;\;\;\;\times \mathbb{P}(S=s) e^{n l s} e^{(1-s) k \sum_{i=1}^n y_{n+i}} \bigg) \\
&= \left( r e^l + (1-r) e^{k^2 + \mu k} \right)^n \left( \frac{e^{ln}}{2} + \frac{1}{2} e^{n (k^2 + \mu k)} \right) ,
\end{split} 
\ee
where in the last line we recall that the first integral has already been calculated in Sect. \ref{subsec:LDPmeans}, whereas the second one can easily be done recognising the i.i.d.\ property of the conditioned process $\mathbf{Y}_n'|\left\lbrace S=0 \right\rbrace$.

The SCGF is obtained as follows:
\be
\begin{split}
\lambda_{M_b}(l,k) &= \lim_{n \ra \infty} \frac{1}{2n} \ln G_{M_b}^r(l,k,2n) \\
&= \frac{1}{2} \ln \left( r e^l + (1-r) e^{k^2 + \mu k} \right) + \lim_{n \ra \infty} \frac{1}{2n} \ln \left( e^{nl} ( 1 + e^{n (k^2 + \mu k - l)} ) \right) \\
&= \begin{cases*}
      \frac{k^2+\mu k}{2} + \frac{1}{2} \ln \left( r e^l + (1-r) e^{k^2+ \mu k} \right) & if $k^2 + \mu k - l > 0$ \\
      \frac{l}{2} + \frac{1}{2} \ln \left( r e^l + (1-r) e^{k^2+ \mu k} \right) & if $k^2 + \mu k - l \leq 0$ .
    \end{cases*} 
\end{split}
\ee
It is analytical everywhere except on the locus of points $k^2 + \mu k - l = 0$ in $\mathbb{R}^2$. The G\"{a}rtner-Ellis theorem can be applied on the differentiable regions, and the convex hull of the large deviation rate function $I_{M_b}(\eta,j)$ can thus be obtained numerically through Legendre-Fenchel transform. Notice here that the function $I_{M_b}(\eta,j)$, as a consequence of Legendre-transforming, presents a flat region of zeros.

\subsubsection{Ratio observable}
\label{subsubsec:RatioLongModel1}

\begin{figure}
    \centering
    \psfrag{a}{$\omega$}
    \psfrag{b}{$I$}
    \psfrag{c}{$\mu=-1.0$}
    \psfrag{d}{$\mu=-0.5$}
    \psfrag{e}{$\mu=0.0$}
    \psfrag{f}{$\mu=0.5$}
    \psfrag{f}{$\mu=1.0$}    
    \subfloat[Ratio rate functions $I_{M_b}(\omega)$]{{\includegraphics[width=5.5cm]{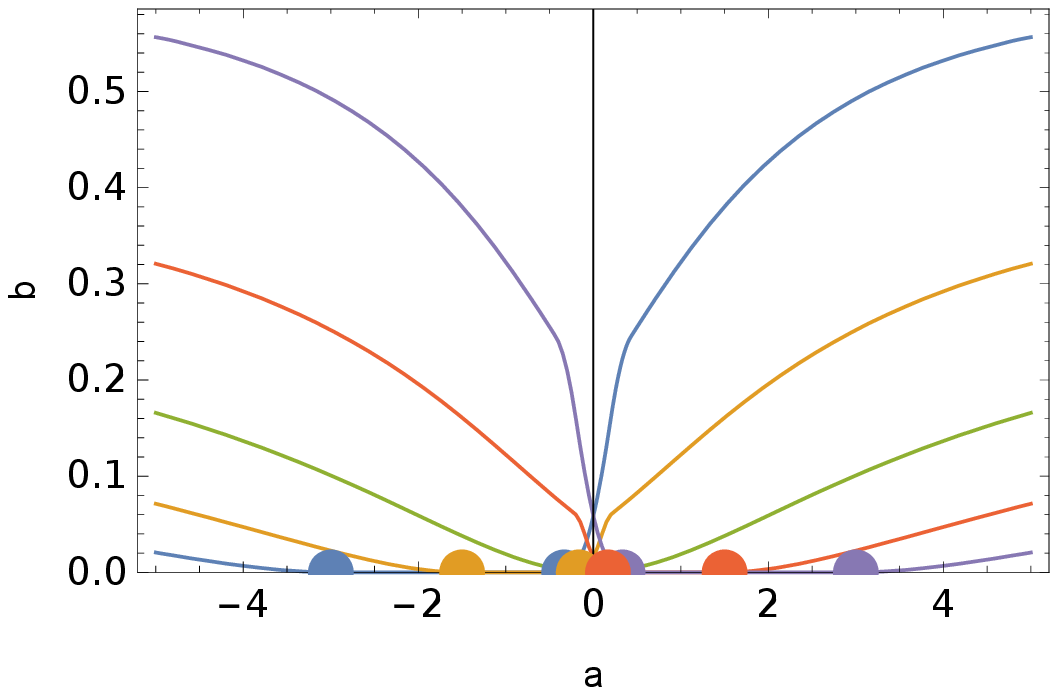}}\label{subfiga:bRatioRate}}%
    \qquad
    \subfloat[Locus of minimizers from contraction ($\mu = -1$)]{{\includegraphics[width=5.5cm]{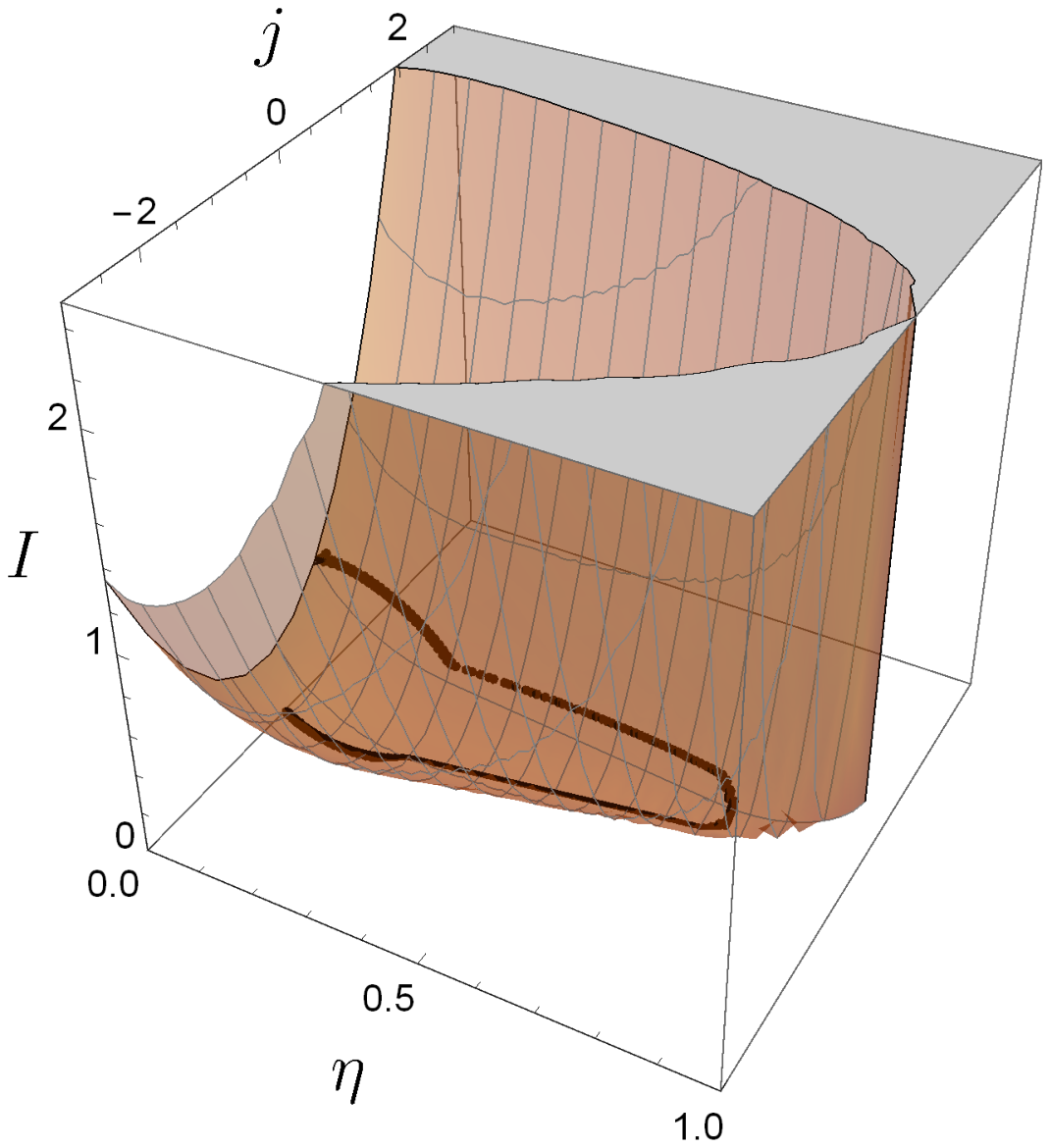} }\label{subfigb:bZerosRatioRate}}%
    \caption{(a) Ratio rate function $I_{M_b}(\omega)$ plotted for $\mu=-1$ (blue), $\mu=-0.5$ (orange), $\mu=0$ (green), $\mu=0.5$ (red), and $\mu=1.0$ (purple). (b) Locus of minimizers $( \eta, \omega \eta )$ satisfying the contraction principle in equation (\ref{eq:LDPratio}) for $\omega \in [-20,20]$ depicted on the surface of the joint rate function $I(\eta,j)$. Numerics obtained for $r=0.5$.}
    \label{fig:bfinal}
\end{figure}

Once again, the large deviation rate function $I_{M_b}(\omega)$ is obtained contracting the joint rate function $I_{M_b}(\eta,j)$ using equation (\ref{eq:LDPratio}). Consistent with the presence of a phase transition in the typical states of the observables $N_n$ and $J_n$, we expect that the observable $\Omega_n$ (for $\mu \neq 0$) has two typical states, also featuring an ergodicity breaking. This is indeed the case, as we can see from Fig. \ref{subfiga:bRatioRate}, where for any curve obtained with $\mu \neq 0$ there is a flat region marking a non-singleton set of zeros. The boundaries of this set are highlighted by coloured dots which mark the two typical states $\hat{\omega}_1 = \hat{j}_1/\hat{\eta}_2 = (1-r)\mu/(1+r)$ and $\hat{\omega}_2 = \hat{j}_2/\hat{\eta}_1 = (2-r)\mu/r$ arising from the ergodicity breaking in the on-off process in the bottom layer.

\begin{figure}
    \centering
    \psfrag{a}{$\omega$}
    \psfrag{b}{$I$}
    \psfrag{c}{\text{theor} $\mu = -1$}
    \psfrag{h}{$n=125$}
    \psfrag{i}{$n=500$}
    \psfrag{l}{$n=1000$}
    \psfrag{m}{$n=2000$}
    \includegraphics[width=0.5\linewidth]{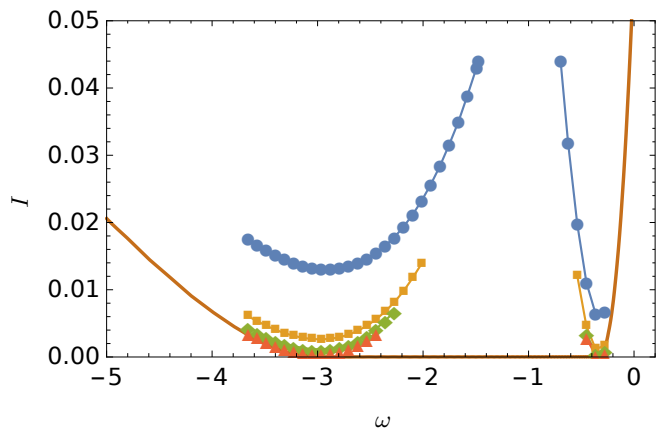}
    \caption{Ratio rate function $I(\omega)$ from theory (solid line) compared with Monte Carlo simulations (symbols) for duration $n = 125,500,1000,2000$ (from top to bottom). Samples of $2 \times 10^7$ trajectories for each simulation.}
    \label{fig:bRatioMonte}
\end{figure}

As evident in Fig. \ref{subfigb:bZerosRatioRate}), the flat region in the ratio rate function corresponds to a set of zeros appearing in the joint rate function $I_{M_b}(\eta,j)$ minimizing the variational equation (\ref{eq:LDPratio}) for $\omega \in \left[\hat{\omega}_1,\hat{\omega}_2\right]$. Notice that this flat region of zeros does not represent a phase coexistence region where fluctuations have a different scaling, as seen for instance in systems like the 2D Ising model \cite{touchette2009large,ellis1995overview}, models of growing clusters \cite{jack2019large}, and critical constrained pinning models \cite{zamparo2019large1,zamparo2019large2}; it is just a consequence of calculating the joint rate function $I_{M_b}(\eta,j)$ through Legendre-Fenchel transform, which gives as output the convex hull of the real joint rate function. To support this argument we compare the ratio rate function $I_{M_b}(\mu)$ obtained for $\mu = -1$ with Monte Carlo simulations in Fig. \ref{fig:bRatioMonte}. Here we see that the simulations, which presumably converge to the true ratio rate function as the trajectory length is increased, do not match with the theoretical curve in the flat region. Instead, they highlight the two typical states and suggests the same scaling of large deviations throughout the domain, indicating once again that the flat part does not constitute a phase coexistence region, but is just the convex hull of the true rate function in that interval.

Although the ratio $\Omega_n$ has an ergodicity-breaking phase transition by construction of the process, tails of the rate functions still seem to be robust. Numerics suggest that the rate function is differentiable, which we believe is a consequence of correlations between the on-off process in the bottom layer and the random walk in the top layer. Indeed, the presence of such correlations gives a curved shape to the joint rate function, and for this reason the locus of minimizers satisfying equation (\ref{eq:LDPratio}) draws a curve without linear sections on $I_{M_b}(\eta,j)$ (see Fig. \ref{subfigb:bZerosRatioRate}). In contrast, model $M_{b1}$ in Appendix B has no correlations between the layers, and shows the appearance of a non-differentiable point at $\omega^* = 0$. A pre-cursor of this non-differentiability can be seen in Fig. \ref{subfiga:bRatioRate} where a rapid change of slope happens close to $\omega = 0$.

Finally, we argue that a flat region in the ratio rate function is manifest generically when a phase transition generates a flat region of zeros (not coincident with the $\eta$ axes) in the joint rate function. The phase transition can be in the bottom layer, as seen here, or directly in the random walk layer, or in both. We also investigated a reset process based on the number of red balls extracted from a Eggenberg-P\'{o}lya urn model with two initial balls: a red one, and a blue one \cite{mahmoud2008polya,feller2008introduction}. In this process the resets are power-law correlated but the ratio rate function is found to be qualitatively equivalent to that shown here, and is not explicitly reported.
 
\section{Conclusion}
\label{sec:conclusions}

We have studied large deviation properties of a ratio observable in stochastic reset processes. We focused on a class of discrete-time processes in which an internal clock (controlled by an on-off process in the bottom layer) restarts with some probability at each step, while a random walk (in the top layer of the model) takes jumps according to a probability density function dependent on the time since reset. In particular, we have shown via contraction, how to derive a large deviation rate function for the ratio observable: current divided by the number of reset steps. Significantly, the large deviation rate function so obtained is non-convex and has tails bounded by horizontal asymptotes, which can be derived analytically from fluctuation properties of the empirical mean current and the empirical mean number of reset steps. We regard the presence of these tails as a universal feature characterising ratios of observables in cases where the denominator can be arbitrary close to $0$. Technically this corresponds to the ratio rate function being weak, which is a signature of the well-known fact that often ratio observables have heavy-tailed distributions. In contrast to the large deviation rate function of the efficiency studied in stochastic thermodynamics, our ratio rate function does not have a maximum. Such a maximum corresponds to a phase transition in the fluctuations of the efficiency and we assert that this is a consequence of having a denominator that can take both positive and negative values, which cannot happen in our case as the number of reset steps must be positive.

We argue that whenever the reset nature of the process is conserved, meaning that the random walk in the top layer is coupled to the bottom on-off process, the ratio large deviation rate function has the general properties described above and is differentiable. We demonstrated this for two particular models with dynamical phase transitions in the current and/or on-off processes. The ratio rate function was found to be robust in the presence of such dynamical phase transitions although, when long-range correlations are present, the convex hull of the rate function has a flat region of zeros instead of a single minimum. The boundaries of this interval represent the two typical states of the ratio surviving in the thermodynamic limit and correspond to an ergodicity breaking. Physically there is no phase coexistence here; the flat section of the rate function is merely an artifact of the Legendre-Fenchel transform. 

Understanding general features of the ratio observable is potentially important for many interdisciplinary applications, e.g. molecular and nano-motors, where correlations may make it difficult to calculate the rate function analytically. In the particular context of our work here, we note that reset dynamics appear in run-and-tumble models (as used to describe bacterial motility) and such processes can manifest a change of scaling \cite{gradenigo2019first}. It would be interesting to see if the ratio observable is affected by this scaling change and similar kinds of phase transition \cite{nickelsen2018anomalous} and, more generally, if one can obtain probabilistic bounds on the rate function. Mathematically, there are also questions related to the existence of a weak large deviation principle \cite{dembo2010large} when one allows the number of reset steps to be zero. There is much scope for future work, both theoretical and applied.

\begin{acknowledgements}
The authors thank Hugo Touchette for carefully reading the manuscript. F.C. is grateful to Giorgio Carugno and Jan Meibohm for interesting discussions, and to Rapha\"{e}l Chetrite for hospitality at Laboratoire J.A. Dieudonn\'e during the last stages of the work. F.C. also thanks the Institute of Mathematics and its Applications for a small grant, while R.J.H. acknowledges an External Fellowship from the London Mathematical Laboratory. Part of this research utilised Queen Mary's Apocrita HPC facility, supported by QMUL Research-IT. [http://doi.org/10.5281/zenodo.438045.]
\end{acknowledgements}

\appendix

\section{Inverse $z$-transform}
\label{app:InvzTransf}

Here we present two methods, one algebraic and one analytic, to numerically calculate the inverse $z$-transform of the function $\tilde{G}^r_{M_a}(l,k,z)$ of Sect. \ref{subsec:Moda}. Both these methods rely on the fact that $G_{M_a}^r(l,k,n)$ is absolutely summable (all poles of $\tilde{G}_{M_a}^r(l,k,z)$ are in the unit circle), or can be rescaled to be such by appropriately remapping all the poles of $\tilde{G}_{M_a}^r(l,k,z)$. 
In the algebraic method we truncate the $z$-transform at the $N$-th term: $\tilde{G}_{M_a}^r(l,k,z_1) \approx \sum_{n=1}^N G_{M_a}^r(l,k,n) z_1^{-n}$. If $N$ is sufficiently large, this is a good approximation for $\tilde{G}_{M_a}^r(l,k,z_1)$, since $z_1^{-n}$ rapidly tends to $0$ as $n$ increases. In order to invert the approximate transform we need to know the value of $\tilde{G}_{M_a}^r(l,k,z_1)$ at, at least, $N$ different points. Considering $m$ different points in the region of convergence of $\tilde{G}_{M_a}^r(l,k,z)$ leads to the approximate system of equations:
\[
\begin{bmatrix}
\tilde{G}_{M_a}^r(l,k,z_1) \\
\tilde{G}_{M_a}^r(l,k,z_2) \\
\vdots \\
\tilde{G}_{M_a}^r(l,k,z_m)
\end{bmatrix}
\approx
\begin{bmatrix}
1 & z_1^{-1} & z_1^{-2} \dots & z_1^{-N} \\
1 & z_2^{-1} & z_2^{-2} \dots & z_2^{-N} \\
\vdots & & \ddots & \vdots \\
1 & z_m^{-1} & z_m^{-2} \dots & z_m^{-N}
\end{bmatrix}
\begin{bmatrix}
G_{M_a}^r(l,k,0) \\
G_{M_a}^r(l,k,1) \\
\vdots \\
G_{M_a}^r(l,k,N)
\end{bmatrix} .
\] 
This can be rewritten as $\mathbf{\tilde{G}} \approx \mathbf{A} \mathbf{G}$. Assuming $m=N$ this system has a unique solution provided that $\mathbf{A}$ is full rank. For our purposes, following \cite{merrikh2014two}, we consider $m$ bigger than $N$, and the solution to $\mathbf{\tilde{G}} \approx \mathbf{A} \mathbf{G}$ is obtained by finding the $\mathbf{G}$ which minimizes $\lVert \mathbf{\tilde{G}} - \mathbf{A} \mathbf{G} \rVert_2$.

The analytic method is based on transforming the $z$-transform into a discrete Fourier transform and then applying well-known routines for calculating its inverse. The first step is to substitute $z=e^{i \omega}$ in $\tilde{G}_{M_a}^r(l,k,z)$, making the latter periodic in $\omega$. Then we obtain a finite sample by taking $\omega = 2 \pi k / M$ for integer $k \in [0,M-1]$, and finally calculate the inverse discrete Fourier transform. Just like the previous method, this works provided that $G_{M_a}^r(l,k,n)$ is absolutely summable, or rescaled to be such, and gives a better approximation for bigger values of $M$.

The plots in Fig. \ref{fig:aCompSCGFInvFour} are obtained using $m=430$, and $M=N=400$.

\section{Infinite-time correlations: $M_{b_1}$}
\label{app:Mb1}

We study here a modified version of the model $M_b$ in Sect. \ref{subsec:Modb}. Here the random walk in the top layer is uncoupled from the bottom on-off process, eliminating the reset nature.
Due to this change there is no need to distinguish $\mathbf{Y}_n$ from $\mathbf{Y}_n'$, and we will write the full top-layer process as $\mathbf{Y}_{2n}$. Specifically, regardless of what happens in the bottom layer, the random walk takes a jump according to a Gaussian distribution of mean $\mu$ and variance $\sigma^2 = 2$. The observable $N_{2n}$ in the bottom layer behaves as already seen in Sect. \ref{subsec:Modb}, whereas the observable $J_{2n}$, being independent from the resets, does not present any DPT. Its rate function is that of a Gaussian random walk characterised by symmetric fluctuations around a single typical value $\mu$.  
Even though the random walk steps are i.i.d., we still expect that the rate function for the observable $\Omega_{2n}$ behaves similarly to that in Sect. \ref{subsubsec:RatioLongModel1}. This is because the ergodicity is broken in the bottom layer, and the presence of two typical states for the observable $N_{2n}$ influences the ratio $J_{2n}/N_{2n}$. In particular, the observable $\Omega_{2n}$ should also have two typical states: $\hat{\omega}_1 = 2 \mu/r$ and $\hat{\omega}_2 = 2 \mu/(1+r)$.

Since the bottom on-off process and the random walk are two independent processes, the joint SCGF can be written as a sum of the SCGFs for the independent observables $N_{2n}$ and $J_{2n}$, i.e. $\lambda_{M_{b_1}}(l,k) = \lambda_{M_{b_1}}(l) + \lambda_{M_{b_1}}(k)$. From the logarithmic moment generating functions, we find
\be
\begin{split}
\lambda_{M_{b_1}}(l) &= 
\begin{cases*}
      \frac{\ln \left( r e^l + (1-r) \right) + l}{2} & if $l > 0$ \\
      \frac{\ln \left( r e^l + (1-r) \right)}{2} & if $l \leq 0$ ,
\end{cases*} \\
\lambda_{M_{b_1}}(k) &= k^2 + \mu k .
\end{split}
\ee
The joint SCGF obtained is analytic everywhere in $\mathbb{R}^2$ except at $l=0$. The joint large deviation rate function $I_{M_{b_1}}(\eta,j)$ can be numerically retrieved through Legendre-Fenchel transform, and from there by contraction we can get $I_{M_{b_1}}(\omega)$.

\begin{figure}
    \centering
    \psfrag{a}{$\omega$}
    \psfrag{b}{$I$}
    \psfrag{c}{$\mu=-1.0$}
    \psfrag{d}{$\mu=-0.5$}
    \psfrag{e}{$\mu=0.0$}
    \psfrag{f}{$\mu=0.5$}
    \psfrag{g}{$\mu=1.0$}
    \captionsetup{font=footnotesize}
    \subfloat[Ratio rate function $I_{M_{b1}}(\omega)$]{{\includegraphics[width=5.5cm]{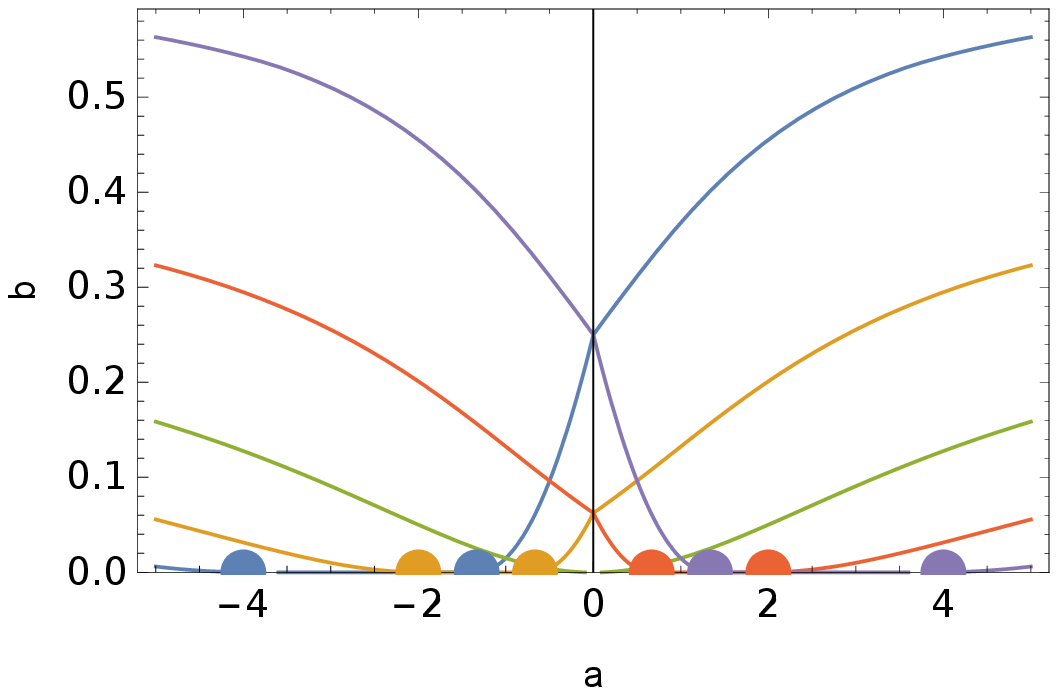}}\label{subfiga:b1RatioRate}}%
    \qquad
    \subfloat[Locus of minimizers from contraction ($\mu = -1$)]{{\includegraphics[width=5.5cm]{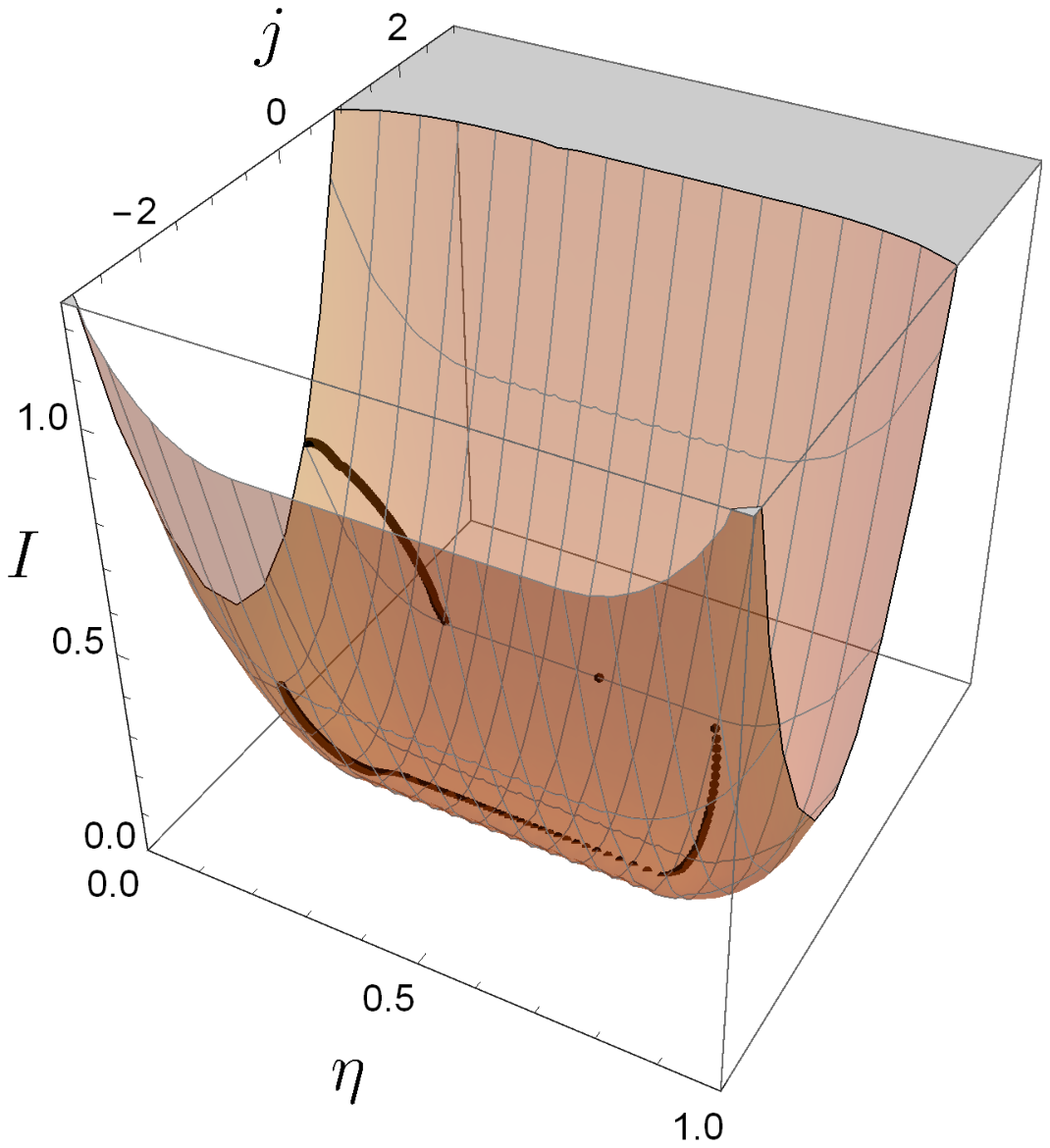} }\label{subfigb:b1ZerosRatioRate}}
    \caption{(a) Ratio rate function $I_{M_{b1}}(\omega)$ plotted for $\mu=-1$ (blue), $\mu=-0.5$ (orange), $\mu=0$ (green), $\mu=0.5$ (red), and $\mu=1.0$ (purple). (b) Locus of minimizers $( \eta, \omega \eta )$ satisfying the contraction principle in equation (\ref{eq:LDPratio}) for $\omega \in [-20,20]$ depicted on the surface of the joint rate function $I_{M_{b1}}(\eta,j)$. Numerics obtained for $r=0.5$.}
    \label{fig:b1final}
\end{figure}

\begin{figure}
    \centering
    \psfrag{a}{$\omega$}
    \psfrag{b}{$I$}
    \psfrag{c}{\text{theor} $\mu = -1$}
    \psfrag{h}{$n=125$}
    \psfrag{i}{$n=500$}
    \psfrag{l}{$n=1000$}
    \psfrag{m}{$n=2000$}
    \captionsetup{font=footnotesize}
    \includegraphics[width=0.5\linewidth]{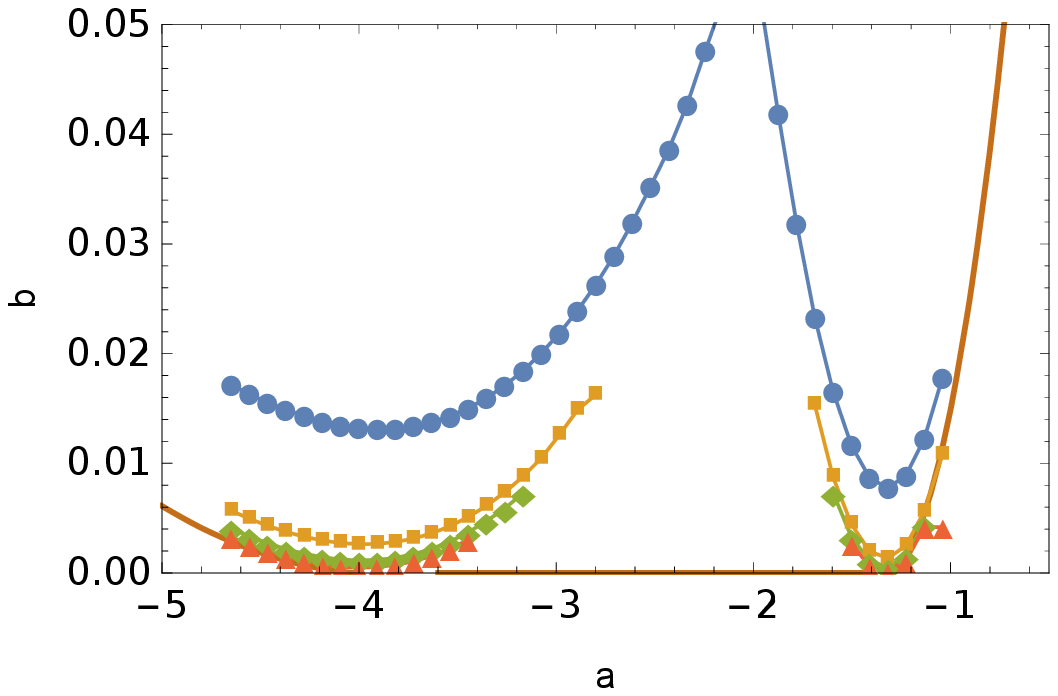}
    \caption{Ratio rate function $I_{M_{b1}}(\omega)$ from theory (solid line) compared with Monte Carlo simulations (symbols) for duration $n = 125,500,1000,2000$ (from top to bottom). Samples of $2 \times 10^7$ trajectories for each simulation.}
    \label{fig:Mb1RatioMonte}
\end{figure}

As expected, $I_{M_{b_1}}(\omega)$ is robust in the tails and presents a flat region of zeros bounded by the typical values $\hat{\omega}_1$ and $\hat{\omega}_2$, see Fig. \ref{subfiga:b1RatioRate}. 
Just like for model $M_b$, we should not confuse this flat region, arising as a natural consequence of Legendre-Fenchel transforming the non-differentiable joint SCGF $\lambda_{M_{b_1}}(l,k)$, with a coexistence phase region. Indeed, fluctuations between the two typical states marked by coloured dots in Fig. \ref{subfiga:b1RatioRate} still scale exponentially linearly in $n$, as indicated by Monte Carlo simulations in Fig. \ref{fig:Mb1RatioMonte}.

\begin{figure}
    \centering
    \psfrag{a}{$\omega$}
    \psfrag{b}{\Large{${'}^I$}}
    \psfrag{c}{$\mu=-1.0$}
    \psfrag{d}{$\mu=0.5$}
    \psfrag{e}{$\mu=0.0$}
    \psfrag{f}{$\mu=0.5$}
    \psfrag{g}{$\mu=1.0$}
    \captionsetup{font=footnotesize}
    \includegraphics[width=0.5\linewidth]{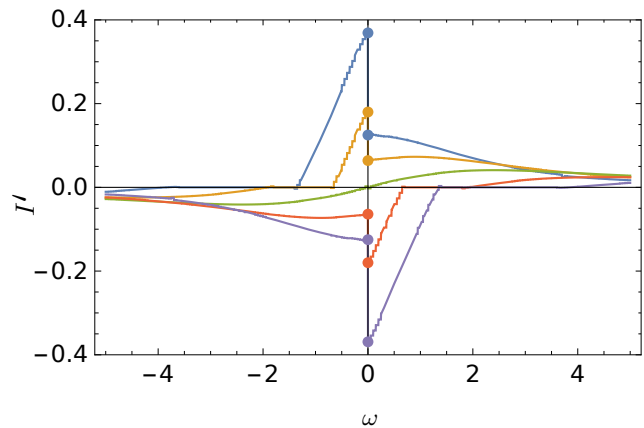}
    \caption{Derivative of ratio rate function $I_{M_{b1}}'(\omega)$ plotted for  $\mu=-1$ (blue), $\mu=-0.5$ (orange), $\mu=0$ (green), $\mu=0.5$ (red), and $\mu=1.0$ (purple). Jumps in the derivative are marked with coloured dots.}%
    \label{fig:b1KinkRatioRate}%
\end{figure}

Although Fig. \ref{subfiga:b1RatioRate} closely resembles Fig. \ref{subfiga:bRatioRate}, one particular feature of the former does not appear in the latter. Uncoupling the random walk in the top layer from the bottom on-off process leads to a genuine `kink' appearing at $\omega^* = 0$ for any $I_{M_{b_1}}(\omega)$ with $\mu \neq 0$. This kink consists of a jump in the first derivative of the function $I_{M_{b_1}}(\omega)$, as evident from Fig. \ref{fig:b1KinkRatioRate} where, for each curve, the left-hand and right-hand limits of the derivative are marked with two dots of the same colour. 
Following Sect. \ref{subsec:RobRate}, it is easy here to prove that $I_{M_{b_1}}(\omega)$ is not differentiable at $0$. We simply need to study $\partial \tilde{I}_{M_{b_1}}(\eta,\omega)/\partial \eta = 0$ and see whether the equation, for $\omega^* = 0$, is verified for more than a single value of $\eta$. The partial derivative reads
\be
\frac{\partial \tilde{I}_{M_{b_1}}(\eta,\omega)}{\partial \eta} =
\begin{cases*}
\ln \frac{-2(1-r)\eta}{r (2 \eta - 1)} + \frac{\omega}{2} (\eta \omega - \mu) & if $\eta \in \left( 0,\frac{r}{2} \right)$ \\
\omega \frac{\eta \omega - \mu}{2} & if $\eta \in \left[ \frac{r}{2},\frac{1+r}{2} \right]$ \\
\ln \frac{(1-r)(1 - 2 \eta)}{2 r (\eta - 1)} + \frac{\omega}{2} (\eta \omega - \mu) & if $\eta \in \left(\frac{1+r}{2},1 \right)$ .
\end{cases*}
\ee
It is clear that for $\omega^* = 0$, the equation is verified for any $\eta \in \left[ r/2,(1+r)/2 \right]$, meaning the solution set of minimizers of (\ref{eq:LDPratio}) for $\omega^*$ is not singleton.

Physically this discontinuity can be considered as a `mode-switching' phase transition in the generation of fluctuations. Moving towards $\omega^*$ from the left we have that fluctuations of the ratio are realized in trajectories with few reset steps, while moving towards $\omega^*$ from the right, they are realized in trajectories having many reset steps. This can also be seen in Fig. \ref{subfigb:b1ZerosRatioRate} where the locus of minimizers $( \eta, \eta \omega )$ shows a numerical jump (independently of the discretization used for $\omega$). This corresponds to the linear section $\eta \in \left[ \frac{r}{2},\frac{1+r}{2} \right]$. In contrast, in model $M_b$ of Sect. \ref{subsec:Modb}, correlations between the bottom process and the random walk in the top layer prevent this sudden switch and no transition of this kind happens. 

\bibliography{mybib}{}
\bibliographystyle{spmpsci}

\end{document}